\documentclass[aps,prl,reprint,superscriptaddress,twocolumn]{revtex4}
\include{math}
\usepackage{bm}
\usepackage{graphicx}
\usepackage{amsmath}

\begin{document}

\title{Magnetic Phase Transitions and Spin Density Distribution in the Molecular Multiferroic GaV$_4$S$_8$ System}

\author{Rebecca L. Dally}
\affiliation{NIST Center for Neutron Research, National Institute of Standards and Technology, Gaithersburg, MD 20899-6102}
\author{William D. Ratcliff II}
\affiliation{NIST Center for Neutron Research, National Institute of Standards and Technology, Gaithersburg, MD 20899-6102}
\affiliation{Department of Materials Science and Engineering, University of Maryland, College Park, MD 20742}
\author{Lunyong Zhang}
\affiliation{Laboratory for Pohang Emergent Materials, Pohang Accelerator Laboratory and Max Plank POSTECH Center for Complex Phase Materials, Pohang University of Science and Technology, Pohang 790-784, Korea}
\affiliation{School of Materials Science and Engineering, Harbin Institute of Technology, Harbin, 150001,  China}
\author{Heung-Sik Kim}
\affiliation{Department of Physics, Kangwon National University, Chuncheon 24341, Republic of Korea}
\affiliation{Department of Physics and Astronomy, Rutgers University, Piscataway, NJ 08854}
\author{Markus Bleuel}
\affiliation{NIST Center for Neutron Research, National Institute of Standards and Technology, Gaithersburg, MD 20899-6102}
\affiliation{Department of Materials Science and Engineering, University of Maryland, College Park, MD 20742}
\author{J. W. Kim}
\affiliation{Department of Physics and Astronomy, Rutgers University, Piscataway, NJ 08854}
\affiliation{Rutgers Center for Emergent Materials, Rutgers University, Piscataway, NJ 08854}
\author{Kristjan Haule}
\affiliation{Department of Physics and Astronomy, Rutgers University, Piscataway, NJ 08854}
\author{David Vanderbilt}
\affiliation{Department of Physics and Astronomy, Rutgers University, Piscataway, NJ 08854}
\author{Sang-Wook Cheong}
\affiliation{Laboratory for Pohang Emergent Materials, Pohang Accelerator Laboratory and Max Plank POSTECH Center for Complex Phase Materials, Pohang University of Science and Technology, Pohang 790-784, Korea}
\affiliation{Department of Physics and Astronomy, Rutgers University, Piscataway, NJ 08854}
\affiliation{Rutgers Center for Emergent Materials, Rutgers University, Piscataway, NJ 08854}
\author{Jeffrey W. Lynn}
\affiliation{NIST Center for Neutron Research, National Institute of Standards and Technology, Gaithersburg, MD 20899-6102}

\date{\today}
\begin{abstract}
We have carried out neutron diffraction and small angle neutron scattering measurements on a high quality single crystal of the cubic lacunar spinel multiferroic, GaV$_4$S$_8$, as a function of magnetic field and temperature to determine the magnetic properties for the single electron that is located on the tetrahedrally coordinated V$_4$ molecular unit. Our results are in good agreement with the structural transition at 44 K from cubic to rhombohedral symmetry where the system becomes a robust ferroelectric, while long range magnetic order develops below 13 K in the form of an incommensurate cycloidal magnetic structure, which can transform into a N{\'e}el-type skyrmion phase in a modest applied magnetic field.  Below 5.9(3) K, the crystal enters a ferromagnetic phase, and we find the magnetic order parameter indicates a long range ordered ground state with an ordered moment of 0.23(1) ${\mu}_\mathrm{B}$ per V ion. Both polarized and unpolarized neutron data in the ferroelectric-paramagnetic phase have been measured to determine the magnetic form factor. The data are consistent with a model of the single spin being uniformly distributed across the V$_4$ molecular unit, rather than residing on the single apical V ion, in substantial agreement with the results of first-principles theory. In the magnetically ordered state, polarized neutron measurements are important since both the cycloidal and ferromagnetic order parameters are clearly coupled to the ferroelectricity, causing the structural peaks to be temperature and field dependent.  For the ferromagnetic ground state, the spins are locked along the $[1,1,1]$ direction by a surprisingly large anisotropy.  
\end{abstract}
\maketitle

\section{Introduction}

The last decade has seen multiferroic materials discovery expand beyond the familiar transition metal oxides to include molecular magnets, \cite{coronado2019molecular, sieklucka2017molecular} where the behaviors of the less localized electrons in molecular multiferroics are the key to their properties. It is therefore vital to understand the concept of molecular orbitals for materials with molecular units. \cite{Launay2013} In contrast, the localized electron picture is the more fundamental concept in the multiferroic transition metal oxides. However, nontrivial physics originating from small molecular-like ion groups, such as the dimer and trimer, have been discovered recently in transition metal oxides. For example, in Fe$_3$O$_4$ \cite{senn2012charge} and Fe$_4$O$_5$, \cite{ovsyannikov2016charge} it's proposed that the collective effects of the electrons are analogous to those in molecular materials, bridging the gap between the two classes of materials. Investigations of the electronic structure in transition metal compounds with molecular-like units is consequently quite interesting and could shed new light on the understanding of novel properties in materials with molecular units.

The molecular magnets in the lacunar spinel family of compounds display properties ranging from multiferroicity to superconductivity under pressure \cite{AbdElmeguid_PRL_2004} to the optical magnetoelectric effect. \cite{okamura2019microwave} Many of the topological phase transitions and quantum properties exhibited are currently of special interest in materials physics research, \cite{BARZ1973983, kim2014spin} including those found in a prototype of this family, GaV$_4$S$_8$. GaV$_4$S$_8$ is cubic ($F\bar{4}3m$) and non-centrosymmetric at ambient temperature, consisting of (V$_4$S$_4$)$^{5+}$ cubane units on a face-centered cubic (fcc) lattice, separated by (GaS$_4$)$^{5-}$ tetrahedra. The four V ions within a cubane unit form a tetrahedron, creating a V$_4$ molecular unit that shares one unpaired electron. \cite{brasen1975magnetic, pocha2000electronic, powell2007cation, widmann2017multiferroic}  At 44 K GaV$_4$S$_8$ undergoes a structural transition to a rhombohedral polar space group ($R3m$) due to a Jahn-Teller distortion where the V$_4$ tetrahedra elongate along $[1,1,1]$-type directions resulting in robust ferroelectricity. The first-order Jahn-Teller transition results in orbitally driven ferroelectricity, leading to GaV$_4$S$_8$'s classification as an improper ferroelectric. \cite{Zhe_PRL_2015} Upon further cooling, long-range incommensurate cycloidal magnetic order develops below 13 K, which is suggested to transform below 6 K into either short- or long-range ferromagnetic order. \cite{white2018direct} Moreover, from the cycloidal phase with the application of modest magnetic fields, a N{\'e}el-type skyrmion lattice forms, which has been investigated in considerable detail using small angle neutron scattering (SANS). \cite{widmann2017multiferroic, kezsmarki2015neel, butykai2017characteristics, ruff2015multiferroicity} GaV$_4$S$_8$, along with VOSe$_2$O$_5$ \cite{Kurumaji_VOSeO_PRL_2017} and fellow lacunar spinel, GaV$_4$Se$_8$, \cite{Bordacs_GVSe_2017} are the only bulk crystals discovered so far to host N{\'e}el-type skyrmion lattices and non-coincidentally, are all polar magnets satisfying the $C_{nv}$ point group symmetry predicted to host such a spin texture. \cite{bogdanov1989therm}

All of the magnetism in GaV$_4$S$_8$ originates from the single unpaired electron shared among the V$_4$ molecular tetramers in the unit cell, \cite{pocha2000electronic, HARRIS19892843} and assessing the distribution of this electron addresses a fundamental question in molecule-based magnetism---as well as in di- and tri-mer physics---concerning the energetic competition and accommodation of the spin and orbital degrees of freedom. \cite{coronado2019molecular, ovsyannikov2016charge, senn2012charge, streltsov2017orbital} For our specific case, in the cubic phase the electron must, by symmetry, be equally shared by the four V ions in the molecule. One of the central unanswered questions, however, concerns the location of the electron in the distorted ferroelectric phase where the apical V shifts and becomes inequivalent to the other three V ions. Related to this problem of the electronic nature of the V$_4$ molecular complex is the question of the coupling of the magnetic order to the ferroelectricity in this multiferroic material and its effect on the cycloidal and skyrmion phases. \cite{cheong2007multiferroics, ratcliff2016magnetic, ratcliff2004multiferroics, nikolaev2020skyrmionic} 

To answer these questions, we have carried out neutron diffraction measurements of the magnetic order as a function of temperature and magnetic field in all the different phases of this multiferroic system, with a particular focus to determine the magnetization density in the ferroelectric phase and the nature of the magnetic ground state, which turns out to be a ferromagnet. The magnetization density, a real-space distribution, is determined by measuring the magnetic form factor, a reciprocal space quantity. This measurement is particularly challenging in GaV$_4$S$_8$ given that the magnetic moment is small ($\approx \frac{1}{4}$ $\mu _\mathrm{B} /\mathrm{V}$) \textit{and} that the magnetic Bragg scattering coincides with the nuclear Bragg scattering. Nevertheless, we have been able to make these measurements by using the field-induced magnetic scattering combined with a polarized neutron beam, which maximizes sensitivity to weak magnetic scattering. 

\section{Methods}
Single crystals were grown via the chemical vapor transport method using a polycrystalline precursor of nominal composition, GaV$_4$S$_8$, and TeCl$_4$ as the transport agent. We note that use of TeCl$_4$ as the transport agent resulted in a higher transport speed when compared to iodine during our synthesis experiments. The polycrystalline precursor was synthesized by a solid state reaction technique where a stoichiometric composition of the elemental powders was heated in an evacuated quartz tube at $100$ $^{\circ} \mathrm{C}$ for two days. Then the tube was heated to $500$ $^{\circ} \mathrm{C}$ at a rate of $5$ $^{\circ} \mathrm{C} / \mathrm{hr}$ and dwelled for four days. Lastly, the tube was furnace cooled to room temperature. The reacted mixture was then well-ground and annealed at $800$ $^{\circ} \mathrm{C}$ for three days, and this annealing procedure was repeated a second time. Then, 1 g of the precursor powder was mixed with about $0.1$ g of TeCl$_4$ and the mixture was sealed in an evacuated quartz tube of 15 cm in length. The powders were placed all at one end of the quartz tube, which was then placed in a tube furnace where the back-growth method was employed to suppress the number of nucleation sites. The temperature gradient profile started with the powder end at $770$ $^{\circ} \mathrm{C}$ and the empty end at $830$ $^{\circ} \mathrm{C}$, which was held for 5 hours. Then, the powder end was heated to $830$ $^{\circ} \mathrm{C}$ at a rate of $5$ $^{\circ} \mathrm{C} / \mathrm{hr}$, and the empty end was cooled to $770$ $^{\circ} \mathrm{C}$ at a rate of $-6$ $^{\circ} \mathrm{C} / \mathrm{hr}$. This final temperature gradient was held for a total growth time of 700 hours, and then the quartz tube was cooled to room temperature at a rate of $100$ $^{\circ} \mathrm{C} / \mathrm{hr}$, and single crystals were obtained where the cold-side of the temperature gradient had been. Sharp magnetic transitions, consistent with literature, were seen in bulk magnetic susceptibility, which was used to initially confirm sample quality (see Supplemental Information \cite{SM}). All the neutron measurements were carried out on one of the larger single crystals weighing 91.1 mg. 

SANS measurements were conducted on the NG-7 SANS instrument primarily using a wavelength of 8 {\AA}, with various guide and detector positions. The sample environment consisted of a closed cycle refrigerator and an electromagnet with holes in the pole pieces to allow the incident neutrons to be approximately parallel to the applied field, which were also parallel to the crystallographic $[1,1,1]$ direction, in order to investigate the cycloidal and skyrmion phases. The wide angle diffraction measurements were conducted on the BT-7 thermal triple-axis instrument \cite{lynn2012double} using a wavelength of 2.359 {\AA} and a closed cycle refrigerator for temperature control.  For the high field data we employed a 10 T cryogen-free magnet.

\begin{figure*}[t]
\includegraphics[scale=0.5]{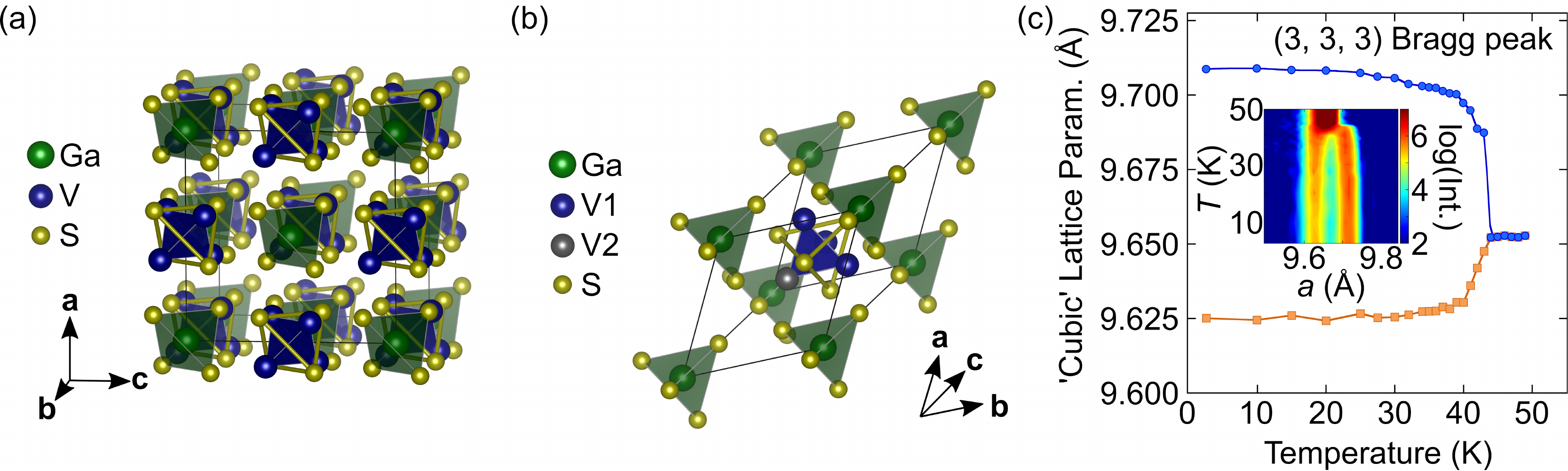}
\caption{(a) Crystal structure of GaV$_4$S$_8$ in the cubic phase and (b) rhombohedral phase, where the apical V2 becomes inequivalent to the three V1 ions. (c) The effective `cubic' lattice parameters from neutron diffraction data. The inset is a contour plot of the neutron diffraction data for the (3, 3, 3) Bragg peak as it splits with temperature below the cubic to rhombohedral phase transition.\label{Fig1}}
\end{figure*}

We also carried-out wide angle polarized neutron measurements, either employing a small ($\approx$ 1 mT) guide field with controllable orientation with respect to the scattering vector, $\mathbf{Q}$, or a 7 T vertical field magnet system. \cite{lynn2012double, chen20073he} The incident beam was polarized with an optically pumped $^3$He cell, with another cell in the scattered beam to analyze the final polarization. The neutron polarization, \textbf{P}, follows the field of the 7 T magnet, however, the actual field that could be applied was restricted to $\leq 2$ T because higher fields quickly depolarized the $^3$He polarizer cell.  Some beam depolarization during the transport through the instrument was present, and the polarizations obtained for these measurements typically provided an instrumental flipping ratio of 15. Two non-spin-flip (NSF) cross-sections, designated ($++$) and ($--$), where the neutron spin does not change orientation, and two spin-flip (SF) cross-sections, ($+-$) and ($-+$), were typically measured.  The field direction for measurements taken in the guide field configuration could be orientated either vertical to the scattering plane (\textbf{Q} $\perp$ \textbf{P}) or along the scattering vector (\textbf{Q} $\parallel$ \textbf{P}), and with no significant transport depolarization, the instrumental flipping ratio was typically 30. The polarization of the cells decreases exponentially with time, and corrections were applied to account for the changes in transmitted polarization and intensity. S{\"o}ller collimations were varied depending on the instrumental resolution requirements.  Most often, we employed coarse resolution in order to integrate over the intensity from the ferroelectric/magnetic domains of each set of `cubic' peaks as discussed below in the text, and consequently, we present the data using cubic notation unless otherwise indicated.  The scattering plane was chosen to be ($H,H,L$), with the cubic lattice parameter of 9.659 {\AA} at low temperatures. Error bars where indicated represent plus and minus one standard deviation of uncertainty.

Density-functional calculations were carried out to compare with the magnetic form factor measurements. For unit cell optimizations (cell volume and shape) and relaxations of initial internal coordinates, the Vienna ab-initio Simulation Package (VASP), which employs the projector-augmented wave (PAW) basis set, \cite{kresse1993ab, kresse1996efficient} was used for density functional theory (DFT) calculations in this work. 330 eV of plane-wave energy cutoff (PREC=high) and $15 \times 15 \times 15$ $\Gamma$-centered $k$-grid sampling were employed. For the treatment of electron correlations within DFT, several exchange-correlation functionals were employed, including Ceperley-Alder (CA) parametrization of local density approximation, \cite{ceperley1980ground} Perdew-Burke-Ernzerhof generalized gradient approximation (PBE) \cite{perdew1996generalized} and its revision for crystalline solids (PBEsol), \cite{csonka2009assessing} DFT+U \cite{dudarev1998electron} on top of LDA, PBE, and PBEsol. $10^{-4}$ eV/{\AA} of force criterion was employed for structural optimizations. 

\section{Results}

We first describe the basic properties of GaV$_4$S$_8$ and characterizations of our single crystal. The sharp transition at 44 K corresponds to the structural transition where the cubic structure distorts and becomes rhombohedral, as shown in Figs.~\ref{Fig1}(a) and (b). This is clearly evident in Fig.~\ref{Fig1}(c) showing the cubic lattice parameter extracted from neutron diffraction measurements of the $(3,3,3)$ Bragg peak. When the peak splits below the Jahn-Teller distortion, two pseudo-cubic lattice parameters can be tracked due to the rhombohedral distortion. This is due to the crystal breaking up into four equally favorable domains, or twins. Below the structural transition temperature, in rhombohedral notation, the peak with a larger ‘cubic’ lattice parameter corresponds to domain 1's $(3, 3, 3)$ lattice plane spacing and the peak with the smaller ‘cubic’ lattice parameter corresponds to three equal rhombohedral lattice plane spacings: domain 2's $(0,0,-3)$, domain 3's $(0, -3, 0)$, and domain 4's $(-3, 0, 0)$ (see Fig.\ S2 for details on the domain definitions). In this distorted phase the four V ions are no longer equivalent, with the apical V2 along the $[1,1,1]$ direction becoming unique compared to the other three, as shown in Fig.~\ref{Fig1}(b). This distortion of the V tetrahedron gives rise to a robust polarization that develops in this improper ferroelectric. With further decrease of temperature, a second transition occurs at 13 K which originates from the development of long range magnetic order as we will discuss in detail below.

\begin{figure}[t]
\includegraphics[scale=0.14]{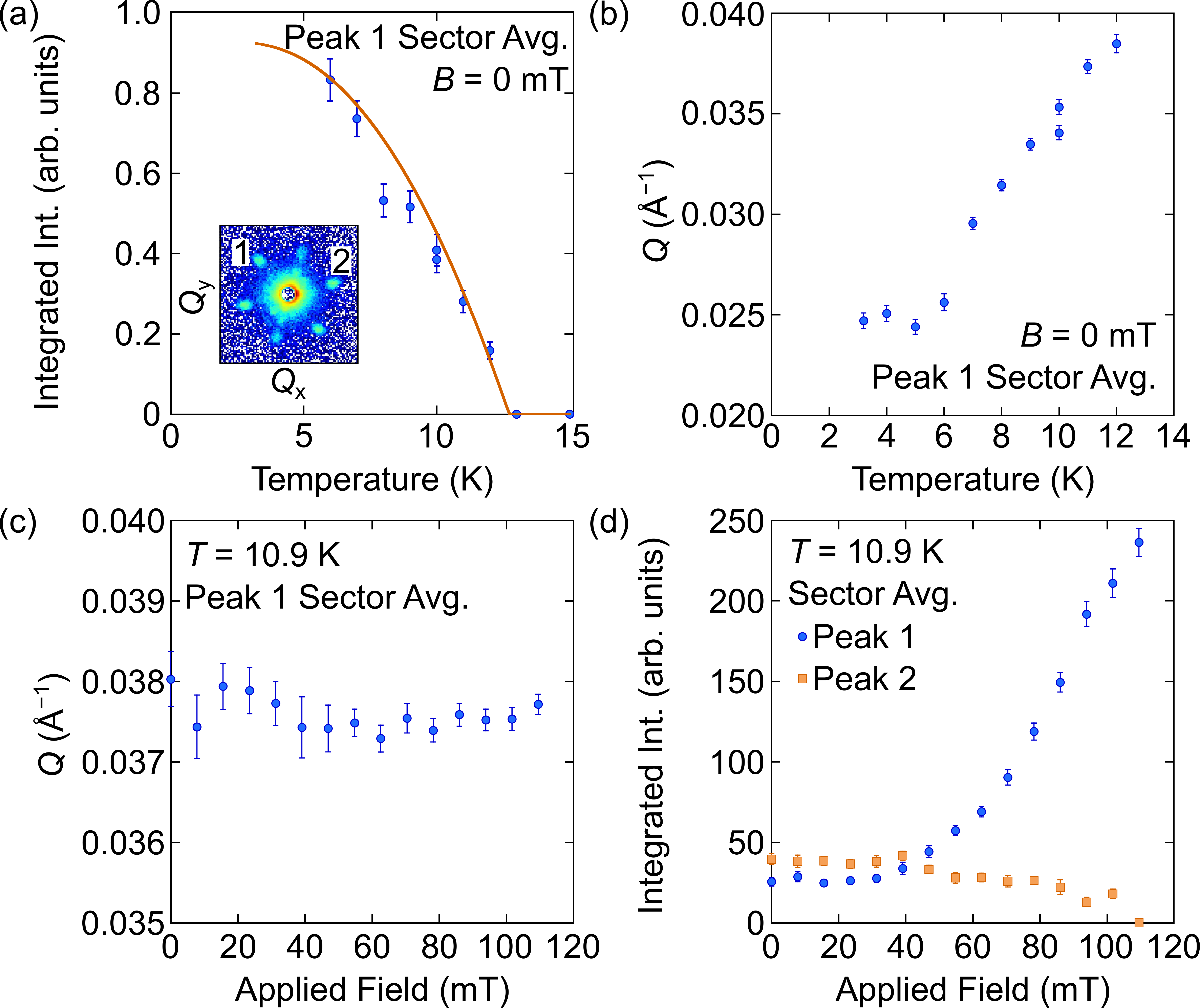}
\caption{Small angle scattering measurements of the wave vector and intensity of the cycloid as a function of temperature and magnetic field.  (a) Integrated intensity of one of the peaks, with the solid curve a simple fit to mean field theory to estimate the ordering temperature of 12.8(3) K. The inset shows a typical example of the observed SANS data, with some background intensity around the beam stop in the center and six diffraction peaks from different domains of the cycloid. The labels, ``1'' and ``2'', are a reference to where the other data in this figure were taken. Sector averages were taken $\pm 11 ^{\circ}$ about the centers of these peaks. (b) Wave vector of the cycloid is strongly temperature dependent, while (c) it is essentially independent of applied magnetic field.  (d)  The intensity, on the other hand, is strongly field dependent.\label{Fig2}}
\end{figure}

To ensure the complex magnetic phase diagram of our sample was consistent with that reported, a series of SANS measurements in temperature-magnetic field phase space were performed and are presented in Fig.~\ref{Fig2}. The inset of Fig.~\ref{Fig2}(a) shows a typical SANS pattern of the sample in the cylcoidal phase. It should be noted that this six-fold pattern resembles SANS data of a skyrmion lattice; careful analysis of the how the cycloid SANS pattern from the four different crystallographic domains cuts through the detector plane has shown why this is the case. \cite{white2018direct} The cylcoidal phase is tracked via the increase in intensity below $T = 12.8(3)$ K, determined by a mean-field fit of the data shown as a solid line in Fig.~\ref{Fig2}(a). At the onset, the cycloid wavevector is $\approx 150$ {\AA} and is strongly temperature dependent as shown in Fig.~\ref{Fig2}(b), and a finite wavevector can be tracked below the nominally reported ferromagnetic transition at $T \approx 6$ K. The field dependence of the wavevector and integrated intensity at $T=10.9$ K are shown in Figs.~\ref{Fig2}(c) and (d), respectively. Unlike the temperature dependence, the wavevector is almost field-independent, at $\approx 0.037$ \AA$^{-1}$ to $0.038$ \AA$^{-1}$ corresponding to a $d$ spacing of 16.8 nm, even as the system passes into the skyrmion phase at $B \approx 30$ mT. Each Bragg peak is a superposition of intensity from a combination of the different crystallograhpic domains. This can be seen by tracking the peak intensity in Fig.~\ref{Fig2}(d). If each peak were Bragg reflections from a single domain, they would behave in the same manner when navigating phase space. 

All of the magnetic scattering originates from a single electron on the V sublattice, as previously mentioned. In the cubic phase (above 44 K), this electron must, by symmetry, occupy a molecular orbital with equal probability on all four V, with the individual orbitals on each ion having basic $3d$ character. The lattice distorts below 44 K, and the ferroelectricity is associated with the displacement of the apical V2 along the $[1,1,1]$ direction, thus making this V inequivalent to the other three V1 ions.  Then the question is whether this crystal distortion substantially changes the probably of finding the electron on the various V ions.  The answer to this question is contained within the magnetic form factor, which is a challenging measurement because of the very small magnetic signal which can be below 1\% of the nuclear intensities. There are three ways the magnetic form factor can be determined. Firstly, one can measure the intensities of a series of magnetic Bragg peaks in the cycloidal phase, and then extract the form factor from those integrated intensities. However, because the wave vector is quite small, such measurements are very difficult unless using SANS, where only the first order peaks were observed (i.e.\ a \textit{series} of peaks could not be measured) as demonstrated in Fig.~\ref{Fig2}. Secondly, one could obtain a series of magnetic Bragg peak measurements in the ground state with high-resolution, wide angle diffraction measurements, but a detailed refinement of the magnetic structure would then need to be carried out, which would require a detailed knowledge of the magnetic and ferroelectric domain populations in the crystal. We deemed that such measurements were not feasible on this system. Thirdly, and the technique that we have chosen to employ, is to apply a large magnetic field to induce all the spins to align. In this case the induced (ferro)magnetic intensities coincide with the structural Bragg peaks. One then can compare the integrated intensities at high field with those in zero field. 

In the paramagnetic state (above 13 K), the zero field data are purely structural, and when using unpolarized neutrons, the magnetic intensity at high field simply adds to the structural intensity for each peak. Then a simple subtraction typically should identify the magnetic component at each Bragg peak, if the structural intensities are field independent, and hence determine the magnetic form factor.  The nuclear structure factors can usually be calculated with good precision and then used as a reference to put the magnetic intensities on an absolute basis. Additionally, when a polarized neutron beam is used, the sensitivity of the form factor measurement is greatly amplified. What follows are results utilizing the strengths of both the unpolarized and polarized neutron beam techniques. For more detailed information on form factor measurements and analysis, see Supplemental Information \cite{SM} and Ch.\ 2 in Ref.~\onlinecite{williams1988polarized}. 

\begin{figure}[t!]
\includegraphics[scale=0.229]{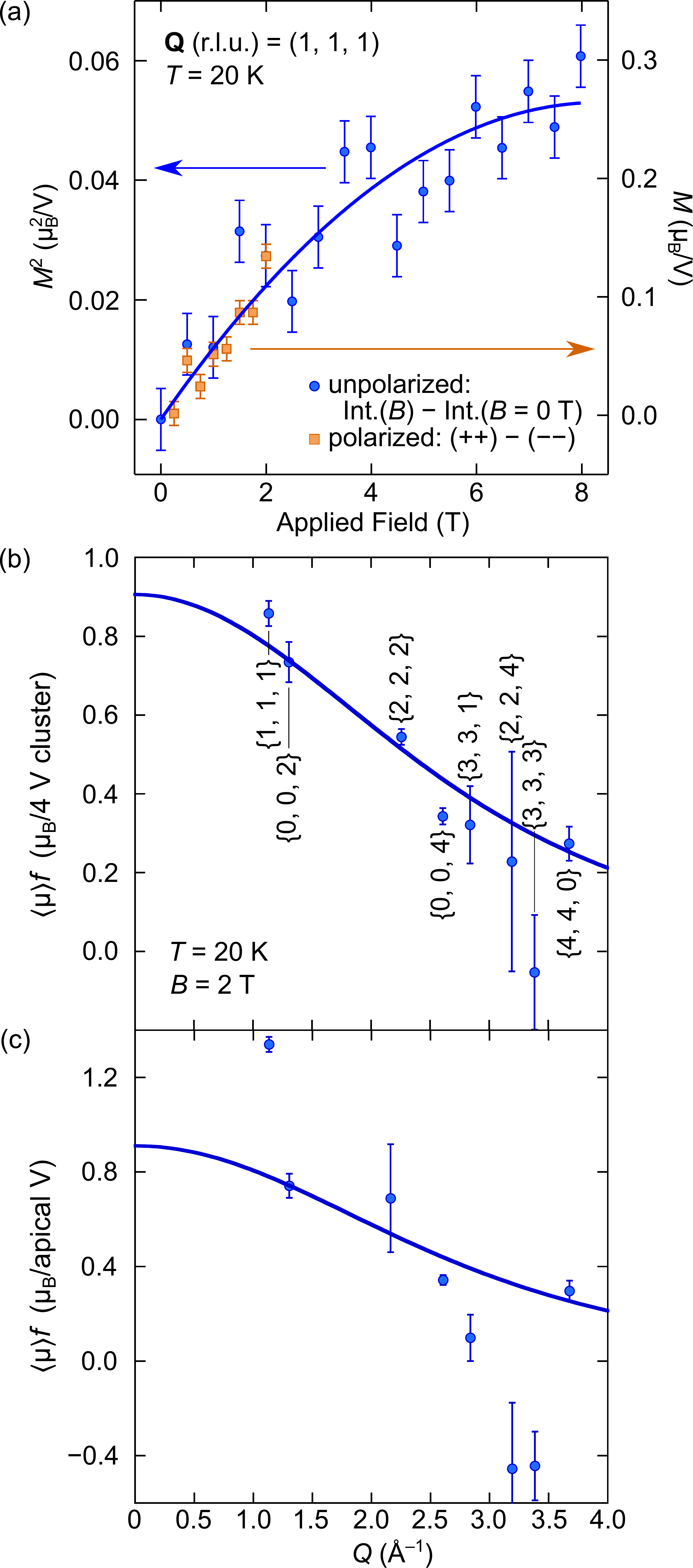}
\caption{(a) Magnetic intensity of the ($1,1,1$)-type Bragg reflections at 20 K. Unpolarized neutron data are plotted against magnetic moment squared per V atom, $M ^2$ ($\mu _B ^2$/V), on the left axis, where the solid line is a guide to the eye. This result was confirmed by employing polarized neutron measurements, shown by the solid squares and plotted against magnetic moment per V atom, $M$ ($\mu _B$/V), up to the maximum field for this technique of 2 T. (b)  Magnetic form factor obtained from the polarized beam data plotted against moment per V$_4$ cluster and using the assumption that the electron is equally likely to be found on the 4 V. The solid curve is the spherical average form factor for a singly ionized V atom. This function was fit to the data, and both data and function are scaled by the saturated moment at high field. (c) Alternative extracted form factor assuming the electron is localized on the apical V, where the solid curve is the spherical average for a singly ionized V atom. This function was fit to the data and both data and function are scaled by the saturated moment at high field.\label{Fig3}}
\end{figure}

The net magnetic integrated intensity, in terms of magnetic moment squared, of the `cubic' $(1,1,1)$-type Bragg reflections is shown in Fig.~\ref{Fig3}(a). Here, coarse instrumental resolution was employed so that the instrument integrated over both ($1,1,1$)-type peaks in the distorted rhombohedral phase to take into account any possible changes in domain populations. These data were taken at 20 K where the ferroelectric order has fully developed, but well above the magnetic ordering temperature of 13 K. The field-induced intensity increases smoothly with increasing field as expected, indicating that the magnetic intensity simply adds to the nuclear intensity. Using a polarized neutron beam, we measured the ($++$) and ($--$) intensities, where the magnetic structure factor either adds or subtracts coherently from the nuclear structure factor. The $(++) - (--)$ difference is then the net magnetic intensity, and we see that the field-dependent data shown by the solid squares in Fig.~\ref{Fig3}(a) are in excellent agreement with the unpolarized beam data where they overlap, demonstrating that the field-induced unpolarized beam measurements are reliable. The magnetic form factor obtained from these polarized beam data are shown in Fig.~\ref{Fig3}(b), where we have assumed in calculating the magnetic structure factor that the average aligned moment on each V ion is the same and aligned along the field direction, albeit not fully saturated at 2 T. The smooth curve is the calculated spherically averaged form factor for a singly ionized V atom. Some deviations from spherical symmetry are not unusual, so we don't expect all the points to be within the statistical uncertainties of the curve, but there is quite good agreement with the overall trend. In contrast, assuming that the unpaired electron is only located on the apical ion produces an unphysical result, as shown in Fig.~\ref{Fig3}(c) for the same experimental data. There could certainly be some modest difference in electron occupancy on the four V ions in the ferroelectric phase as suggested by the DFT calculations we will discuss, but the conclusion is that the average moment on each V is close to being the same. We also note that we attempted to take measurements above the Jahn-Teller distortion at $T=50$ K, however thermal fluctuations were too strong to sufficiently polarize the spins. 

\begin{figure}[t]
\includegraphics[scale=0.16]{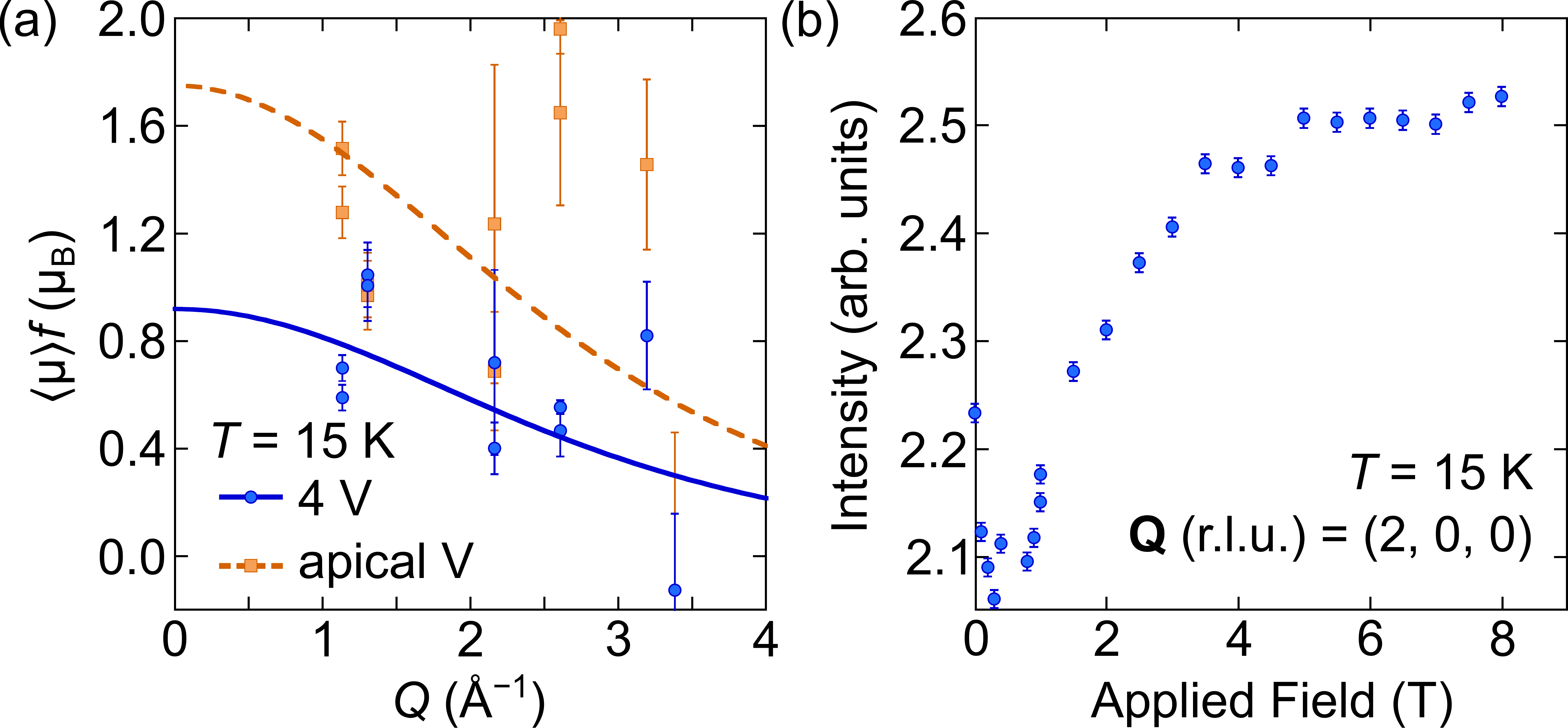}
\caption{(a) Magnetic form factor determined at 15 K, showing that the assumption of all 4 V having approximately the same moment provides a substantially better description of the data than the assumption that all the moment resides solely on the apical V. The solid curve is the spherical average form factor for a singly ionized V atom. This function was fit to the 4 V modeled data, and both data and function are scaled to match the saturated moment at high field. The dashed curve is also the spherical average form factor for a singly ionized V atom but was fit to the apical only model. This curve and data were scaled by the same factor as for the 4 V data.  (b) Field-induced intensity for the ($2, 0, 0$) peak at 15 K, just above the cycloidal ordering temperature.  The intensity first decreases with field, indicating that the structural intensity is field dependent at this temperature.\label{Fig4}}
\end{figure}

We now address the behavior of the system near and below the magnetic ordering temperature(s). For the first set of field-induced data we chose a temperature of 15 K, close to, but above the ordered state where the susceptibility would be large providing an optimal magnetic signal. A close examination of the pyroelectric measurements from Ref.~\cite{ruff2015multiferroicity} shows that the ferroelectric polarization starts to change as the cycloidal magnetic order begins to develop.  The field-induced data at 15 K are shown in Fig.~\ref{Fig4}(a), with the form factor extracted using the two different assumptions: equal moments on the 4 V, and all the moment on the apical V.  We see that the equal-moment assumption provides the better description of the data.  However, we also noticed that there were some anomalies in the intensity versus field curves we obtained, as indicated in Fig.~\ref{Fig4}(b) for the (2, 0, 0) peak.  Here we see that the intensity first exhibits a small decrease with field before increasing. Because the structural and magnetic intensities simply add for unpolarized neutrons, the intensity must monotonically increase with field unless the structural intensity also changes. Consequently, we must conclude that the magnetism exhibits a significant coupling to the ferroelectric order parameter in this temperature regime, even though long range order has not yet developed. Fortunately, the changes in the structural peaks at this temperature are relatively small, and consequently we still were able to extract a reasonable magnetic form factor, but this coupling does introduce some additional systematic uncertainties for the form factor.  The overall conclusion, nevertheless, is that the electron is distributed equally on all four V to a reasonable approximation.

\begin{figure*}[t]
\includegraphics[scale=0.30]{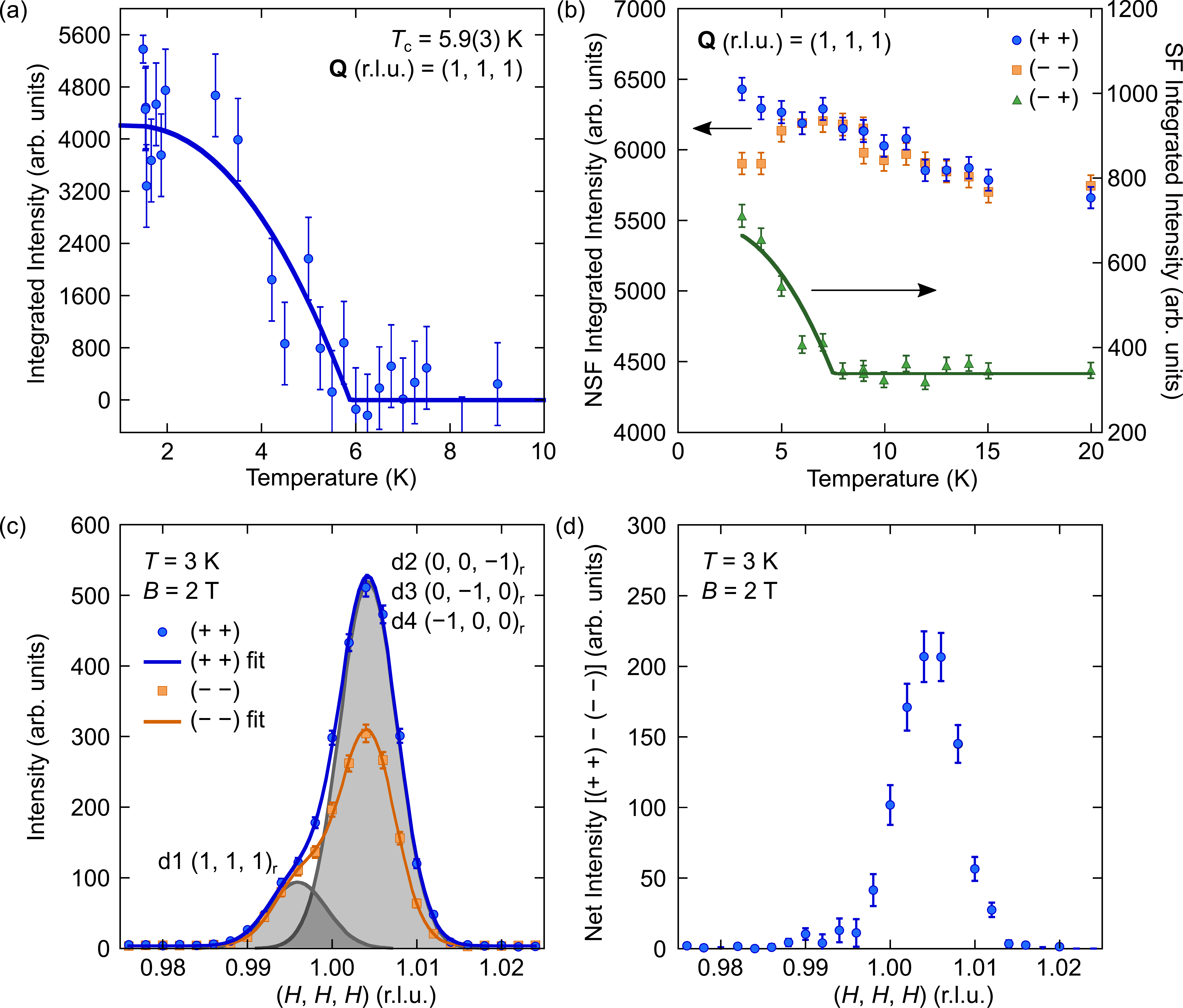}
\caption{(a) Temperature dependence of the integrated intensity of the ($1,1,1$) peak.  The solid curve is a fit to mean field theory to estimate the Curie temperature of 5.9 K.  (b) Polarized beam data of the ($1,1,1$) peak showing that a splitting of the ($++$) and ($--$) scattering at the ferromagnetic transition.  The spin-flip scattering shows that the beam is becoming depolarized below the transition with just a guide field applied to the sample.  (c, d) Ground state (3 K) polarized beam data with high instrumental resolution for the cubic ($1,1,1$) peak, which splits into the $(1,1,1)_r$ peak from rhombohedral domain 1 (d1) and the $(0,0,-1)_r$, $(0,-1,0)_r$, and $(-1,0,0)_r$ peaks from rhombohedral domains 2, 3, and 4 (d2--d4), respectively. Essentially all the magnetic intensity occurs on the d2--d4 peaks even with a 2 T vertical field applied, indicating that the magnetic anisotropy is large enough that the field is insufficient to rotate the spins fully along the vertically applied field direction. \label{Fig5}}
\end{figure*}

With further decrease of temperature the long wavelength incommensurate structure locks into a ferromagnetic state.  Fig.~\ref{Fig5}(a) shows the intensity of the ($1,1,1$) peaks, again taken with coarse resolution so that both types of rhombohedrally distorted peaks are measured simultaneously. The solid curve is a simple fit to mean field theory to estimate an ordering temperature of 5.9(3) K.  The ordered moment obtained from the 3 K data is 0.94(4) ${\mu}_B$/formula unit, or 0.23(1) ${\mu}_{\mathrm{B}}$/V assuming the moment is equally distributed.  This is the same value of the moment as the saturated moment induced at high fields, indicating that the ferromagnetic structure is collinear to a good approximation, as expected from the DFT calculation.  Fig.~\ref{Fig5}(b) shows the polarized beam intensity using a vertical (1 mT) guide field, where we see a difference in scattering between ($++$) and ($--$) below the ordering temperature, which confirms that a ferromagnetic component has developed.  We also see an intensity increase in the spin-flip scattering, which indicates that the development of domains in the ferromagnetic state begins to depolarize the beam.  The conclusion is that the order that has developed is long range in nature.

The data in Fig.~\ref{Fig3}(a) show that well above the magnetic ordering temperature the coupling between the magnetism and structure is not significant, where the magnetic intensity increases monotonically with applied field and quantitatively agrees with the polarized beam data. To investigate the magnetic form factor in the ferromagnetic ground state we needed to apply the polarized beam technique at low temperatures (3 K), since we know there is significant coupling to the lattice, especially considering the very small magnetic signal compared to the structural intensities. The polarized beam technique is required because it unambiguously separates the magnetic and structural cross sections. 

We first collected integrated intensity data using only a guide field, but as Fig.~\ref{Fig5}(b) shows, the beam becomes partially depolarized, and it was found that the depolarization was dependent on how the sample was cooled in the guide field.  With an unknown distribution of domains, it was not possible to extract a reliable form factor. Alternatively, we applied a 2 T vertical field with polarized neutrons and collected a series of Bragg peak intensities, as we did for the 20 K data.  However, we still encountered considerable difficulty in extracting a reasonable form factor.  To investigate possible reasons for these difficulties, we collected time-consuming data employing the highest instrumental resolution available on the instrument. The high instrumental resolution polarized beam data for the cubic $(1,1,1)$ peak, which separates into the $(1,1,1)_r$ peak from rhombohedral domain 1 (d1) and the $(0,0,-1)_r$, $(0,-1,0)_r$, and $(-1,0,0)_r$ peaks from rhombohedral domains 2, 3, and 4 (d2--d4), respectively, is shown in Fig.~5(c). A schematic of the domain definitions can be found in the Supplemental Information. \cite{SM} Surprisingly, we see that essentially all the magnetic intensity occurs on the d2--d4 peak with a 2 T vertical field applied, with very little intensity on the ($1,1,1$)$_r$ peak. This intensity disparity is readily observed in the $[(++)-(--)]$ difference plot in Fig.~\ref{Fig5}(d). This is not the expected result if 2 T were sufficient to pull the spins along the vertical applied field direction. If the spins were aligned along the applied field direction, and assuming an equal distribution of the electron between the V$_4$ cluster, the flipping ratio for the ($1,1,1$)$_r$ peak would be $0.50$ (flipping ratio values expected for additional Bragg peaks are presented in Supplemental Information \cite{SM}), whereas the flipping ratio should be 1 if the moments are still along the domains' local $[1,1,1]$ direction. This indicates that in the ferromagnetic state the magnetic anisotropy is larger than expected---large enough that the applied field of 2 T is insufficient to rotate the spins and align them along the field direction, resulting in an unknown spin direction. The uniaxial magnetic anisotropy has been observed in several types of measurements as it affects a variety of magnetic properties. \cite{ehlers2017exchange, okamura2019microwave, padmanabhan2019optically, muller2006magnetic} This ambiguity negated us from quantitatively determining the form factor in the ground state.

\section{Discussion}

A high-quality and well-characterized sample has been studied, where SANS measurements confirm the onset of cycloidal magnetic ordering, and show that the wavevector is strongly temperature dependent, decreasing as thermal fluctuations are minimized and the cycloidal state becomes unstable. There has been an open question about whether ferromagnetic order in the ground state is truly long-range and collinear \cite{widmann2017multiferroic} due to coexistence of the cycloidal and ferromagnetic phases down to as low as 3 K. \cite{clements2019robust} In agreement with those studies, our SANS data also show that the cycloidal and ferromagnetic regime do coexist in a narrow temperature window, however the wide-angle diffraction data show that the high-field saturated moment at $T = 20$ K is the same as the value obtained at $T = 3$ K. Therefore, our conclusion is that GaV$_4$S$_8$ is a collinear ferromagnet in the ground state, driven by a relatively strong uniaxial anisotropy \cite{kezsmarki2015neel} and strong exchange coupling along interplane bonds. \cite{Zhang2017}

The results for the magnetic form factor in the ferroelectric state show no evidence that there is a substantial difference in the electron distribution within the V$_4$ cluster. We have assumed---whether using the 4 V model or the apical only model---that the unpaired electron resides in a $d$-character orbital, which is why the extracted $\left< \mu \right> f$ values in Figs.~\ref{Fig3}(b) and (c) are compared to the spherical distribution form factor for a singly ionized V atom. While intuitively one might expect the the electron density cloud for a molecular unit sharing an electron to encompass the entire V$_4$ cluster, our DFT calculations point to a more localized picture (Fig. S5), which shows the spin density to be distributed on each of the V atoms for the 4 V model, and positioned on the apical V for the apical only model. 

The experimental observations can be compared to results from first-principles electronic structural calculations. DFT calculations were performed using all three exchange-correlation functionals, CA, PBE and PBEsol, with the choice of correlation parameter $U_{eff} \leq 2$ eV, which leads to the prediction of a ferromagnetic ground state with a rhombohedral distortion ($\alpha \approx 59.4^{\circ} $). Note that all results show the distribution of the spin moments across all V atoms in a V$_4$ cluster; 0.419 ${\mu}_{\mathrm{B}}$ and 0.221 ${\mu}_{\mathrm{B}}$ for the apical and other three V sites from PBEsol+Ueff = 2 eV result, respectively. We comment that more elaborate treatment of electron correlations may be required, as discussed in a recent dynamical mean-field study of GaV$_4$S$_8$, \cite{kim2018molecular} where the inclusion of correlations promotes the high-spin nature of the V$_4$ cluster and makes the spin moments even more equally distributed. The inhomogeneous distribution of the electron suggested by DFT does not greatly change the polarized neutron flipping ratio values when compared to the equal distribution model at small moments (as is the case at 20 K and 2 T). These calculations and results can be found in the Supplemental Information (see Ref.~\onlinecite{SM} for further discussion and comparison to the recent study, Ref.~\onlinecite{Wang2019}), and are the justification for analyzing the data with the equal distribution model and apical only model. 

One appealing property of GaV$_4$S$_8$, and other members of the lacunar spinel family, is that they are molecular magnets where intermolecular interactions play the dominant role in the ordered phases despite large distances between clusters. The non-collinearity leading to the cycloid and skyrmion phases in GaV$_4$S$_8$ is understood to be a result of Dzyaloshinskii-Moriya (DM) interaction terms between the molecular $S=1/2$ moments in V$_4$ clusters. The stability of these phases with respect to the spin distribution within a cluster has not been reported, though, and it's expected that the distribution would likely tune the size of the DM interaction, but would not alter the fact that the DM term exists due to the lack of inversion symmetry between clusters. Recently, exchange parameters were estimated from the DFT results of Ref.~\cite{nikolaev2020skyrmionic}, although the calculations used a different formalism than those presented here. 

The molecular orbital configuration for the V$_4$ clusters in the cubic phase contains 7 $3d$ electrons with one electron in a triply degenerate $t_2$ orbital. This orbital degeneracy leads to the Jahn-Teller distortion and the splitting of the $t_2$ band into a lower level $a_1$ band and higher level $e$ band. \cite{pocha2000electronic} Structural instabilities and electrical resistivity changes are common themes in molecular magnets due to the molecular orbital electron configurations, which differ greatly from the localized electron picture. Fe$_4$O$_5$ is another example, where competing dimeric and trimeric orderings lead to a modulated nuclear structure and large increase in electrical resistivity. \cite{ovsyannikov2016charge} 

Electron-electron interactions in molecular magnets are often highly correlated, and in GaV$_4$S$_8$, this correlation occurs within a V$_4$ tetrahedron, but due to larger intercluster distances ($> 4$ \AA), the electrons are localized to the clusters, forming a Wigner glass \cite{Sahoo1993, Rastogi1996} and leading the bulk to a Mott insulating state. The electron configuration of the clusters has been calculated, \cite{muller2006magnetic, pocha2000electronic} and it was shown that the number of electrons available for metal-metal bonding is important in the determination of the Jahn-Teller distortion (elongation versus compression) and that orbital overlap, of primarily $d_z^2$ character, between the apical and basal V ions depends on the rhombohedral angle and is sensitive to small changes in the metal-metal bonding distances. It's possible that these distances can continue to evolve as temperature decreases from the ferroelectric transition. Additionally, it was recently shown that concommitant with the onset of magnetic order, the variable range hopping mechanism changes from Mott to Efros-Shklovskii, \cite{Lal2018} providing additional motivation to tackle the spin-density measurement in the magnetic ground state.
 
The initial onset of polarization below the Jahn-Teller distortion is due to an orbital order mechanism, which remains the dominant source of ferroelectricity throughout the phase diagram. Exchange-striction is the main spin-driven mechanism for ferroelectricity in all of the magnetically ordered phases \cite{ruff2015multiferroicity, Zhang2017}, and is additive to the much larger polarization from the orbital order mechanism. However, even the small spin-driven polarization reflective of the interplay between the electronic and magnetic properties is evident from pyroelectric current data, which show an increase in ferroelectric polarization just above the cycloidal magnetic phase transition, and then a drop just above the ferromagnetic phase transition. \cite{ruff2015multiferroicity} In fact, our neutron data at 15 K demonstrate this interplay. GaV$_4$S$_8$ is still paramagnetic at 15 K, but cycloidal fluctuations due to it's proximity to the magnetically ordered phase contribute to the magnetoelectric coupling as evidenced by the non-monotonic field dependence of the $(2, 0, 0)$ Bragg peak intensity shown in Fig.~\ref{Fig4}(b). Additionally, our $T = 15$ K form factor measurements, to a reasonable degree, still imply that the electron is distributed between the 4 V atoms. 

The fate of the electron's detailed distribution remains an open question in the ferromagnetic ground state. Therefore, in addition to our attempts at measuring the $T=3$ K form factor with a 2 T polarizing field, we also attempted high-resolution polarized neutron measurements with just the application of the small 1 mT guide field, where we could reasonably assume the moments for each domain point along the local $[1,1,1]$-like axes. However, we discovered that in addition to peak splitting due to the Jahn-Teller distortion, there is additional peak splitting---although reversible with temperature---along the sample mosaic direction due to the rhombohedral domains forming separate crystallites (see SFig.\ 1 in Supplemental Information for more detail \cite{SM}). Form factor analysis is dependent on knowing the direction of the spins which contribute to each Bragg peak, and this additional splitting greatly complicated attempts at analysis. In the future, this problem might be overcome if a single-domain sample of sufficient mass could be obtained below the Jahn-Teller distortion. A natural choice is to pole samples with an electric field through the Jahn-Teller distortion to favor a single ferroelectric (and crystallographic) domain, however, GaV$_4$S$_8$ can be a leaky ferroelectric, making this scenario difficult on a sufficient size crystal.

\section{Concluding Remarks} 

Electron sharing and strong correlations in molecular units often lead to exotic orbital hybridization effects and electron configurations. This is why molecular multiferroics are a promising class of materials for which to search for the elusive and simultaneous ordering of magnetism and ferroelectricity, and by researching such compounds, we can leverage what's learned to chemically design new materials. With this motivation in mind, we have studied single crystals of the molecular multiferroic, GaV$_4$S$_8$. The samples were grown using the chemical vapor transport technique and characterization was performed via small-angle neutron scattering measurements, which confirmed sample quality matching that of previous research. The Bragg intensities for a series of peaks in the ferroelectric-paramagnetic phase have been measured with both polarized and unpolarized neutrons to determine the magnetic form factor, and the intensities are consistent with a model of the single spin being approximately uniformly distributed across the V$_4$ molecular unit, rather than residing on the single apical V ion. The experimental data were accompanied with DFT calculations, which also suggested a distribution of the electron density throughout the cluster. Further, via unpolarized and polarized neutron diffraction measurements, we have determined that the magnetic ground state of GaV$_4$S$_8$ is consistent with a collinear ferromagnet.

\begin{acknowledgments}
We are pleased to acknowledge helpful discussions with Daniel Khomskii and Jamie Manson. D.V., H.-S.K., K.H., and S.-W.C.\ are supported by NSF Grant DMREF DMR-1629059. The work at Postech was supported by the National Research Foundation of Korea (NRF) funded by the Ministry of Science and ICT(No. 2016K1A4A4A01922028). The identification of any commercial product or trade name does not imply endorsement or recommendation by the National Institute of Standards and Technology.
\end{acknowledgments}

\bibliography{ms}

\end{document}


\title{Supplemental Information: Magnetic Phase Transitions and Spin Density Distribution in the Molecular Multiferroic GaV$_4$S$_8$ System}

\author{Rebecca L. Dally}
\affiliation{NIST Center for Neutron Research, National Institute of Standards and Technology, Gaithersburg, MD 20899-6102}
\author{William D. Ratcliff II}
\affiliation{NIST Center for Neutron Research, National Institute of Standards and Technology, Gaithersburg, MD 20899-6102}
\affiliation{Department of Materials Science and Engineering, University of Maryland, College Park, MD 20742}
\author{Lunyong Zhang}
\affiliation{Laboratory for Pohang Emergent Materials, Pohang Accelerator Laboratory and Max Plank POSTECH Center for Complex Phase Materials, Pohang University of Science and Technology, Pohang 790-784, Korea}
\affiliation{School of Materials Science and Engineering, Harbin Institute of Technology, Harbin, 150001,  China}
\author{Heung-Sik Kim}
\affiliation{Department of Physics, Kangwon National University, Chuncheon 24341, Republic of Korea}
\affiliation{Department of Physics and Astronomy, Rutgers University, Piscataway, NJ 08854}
\author{Markus Bleuel}
\affiliation{NIST Center for Neutron Research, National Institute of Standards and Technology, Gaithersburg, MD 20899-6102}
\affiliation{Department of Materials Science and Engineering, University of Maryland, College Park, MD 20742}
\author{J. W. Kim}
\affiliation{Department of Physics and Astronomy, Rutgers University, Piscataway, NJ 08854}
\affiliation{Rutgers Center for Emergent Materials, Rutgers University, Piscataway, NJ 08854}
\author{Kristjan Haule}
\affiliation{Department of Physics and Astronomy, Rutgers University, Piscataway, NJ 08854}
\author{David Vanderbilt}
\affiliation{Department of Physics and Astronomy, Rutgers University, Piscataway, NJ 08854}
\author{Sang-Wook Cheong}
\affiliation{Laboratory for Pohang Emergent Materials, Pohang Accelerator Laboratory and Max Plank POSTECH Center for Complex Phase Materials, Pohang University of Science and Technology, Pohang 790-784, Korea}
\affiliation{Department of Physics and Astronomy, Rutgers University, Piscataway, NJ 08854}
\affiliation{Rutgers Center for Emergent Materials, Rutgers University, Piscataway, NJ 08854}
\author{Jeffrey W. Lynn}
\affiliation{NIST Center for Neutron Research, National Institute of Standards and Technology, Gaithersburg, MD 20899-6102}

\date{\today}

\maketitle

\section{Bulk Susceptibility}

\begin{figure}[b]
\includegraphics[scale=0.3]{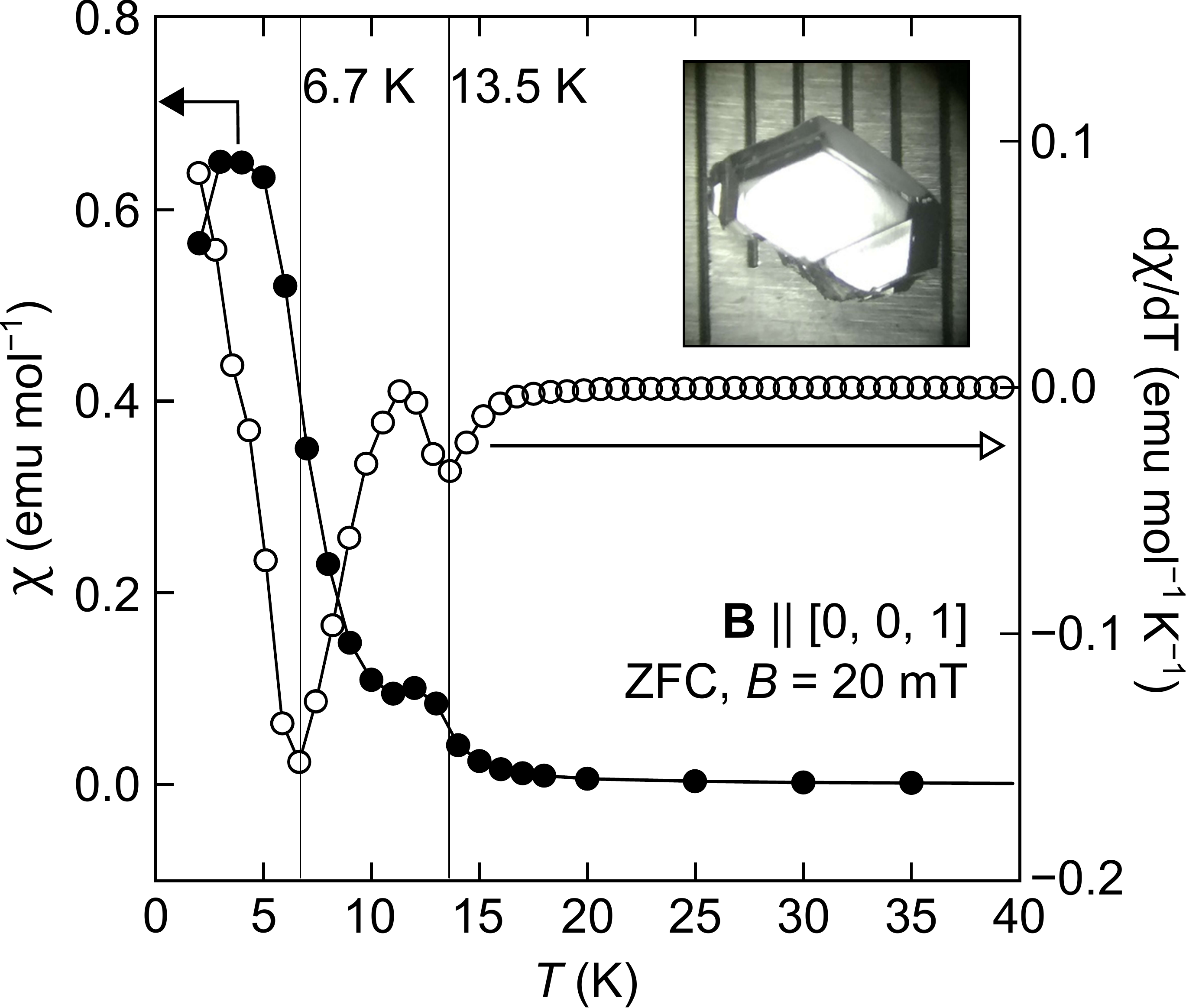}
\caption{Zero-field cooled (ZFC) bulk susceptibility measurements on the crystal used throughout the study, which is shown in the inset and was massed at 91.1 mg. The data show well-defined magnetic transitions at $T=13.5$ K and $T=6.7$ K, corresponding to cycloidal long-range order and ferromagnetic order, respectively. The transitions are even more pronounced in the susceptibility derivative, $d \chi / dT$, shown on the right axis. Note: $1$ emu $= 10^{-3}$ A m$^{2}$.\label{SFig1}}
\end{figure}

Fig.\ S\ref{SFig1} shows the 91.1 mg single crystal sample used throughout this study, along with the zero-field cooled (ZFC) bulk susceptibility measurements used to initially characterize the magnetic transition temperatures. The data show well-defined magnetic transitions at $T=13.5$ K and $T=6.7$ K, corresponding to cycloidal long-range order and ferromagnetic order, respectively, and consistent with previous research.~\cite{Kezsmarki2015} The transitions are even more pronounced in the susceptibility derivative, $d \chi / dT$, shown on the right axis of Fig.\ S\ref{SFig1}. 

\section{Form Factor Calculations}
The single unpaired electron within a V$_4$ molecular cluster is responsible for the magnetic and ferroelectric properties in GaV$_4$S$_8$. Knowing the charge distribution for this unpaired electron within the V$_4$ cluster will guide calculations and understanding of similar systems, and this is the basis for performing the form factor measurements presented in this manuscript. The unpaired electron has a spin, and because neutrons have a magnetic moment, they will interact with unpaired electrons via the dipole-dipole interaction. This contrasts with x-ray scattering (off resonance), where x-rays will interact with all of the electrons in a system, not just those unpaired. 

The structure factor for magnetic Bragg scattering, $F_{\mathrm{M}}(\mathbf{Q})$,  includes the term, $ \mu f(\mathbf{Q})$, which is the magnetic moment multiplied by the magnetic form factor, and $\mathbf{Q}$ is the scattering vector. The magnetic form factor is the Fourier transform of the real-space magnetization density and is responsible for the decrease in magnetic intensity with increasing $Q$. The magnetic structure factor can be written as,

\begin{equation}\label{eq:SFmag}
F_{\mathrm{M}}(\mathbf{Q}) \propto \sum_i \mu _i f_i (\mathbf{Q}) \mathbf{P} \cdot \left[ \widehat{\mathbf{Q}} \left( \widehat{\mathbf{Q}} \cdot \widehat{\bm{ \mu }} _i \right) - \widehat{\bm{ \mu }}_i\right] \exp{(i \mathbf{Q} \cdot \mathbf{R}_i )}
\end{equation}

where $\mathbf{P}$ is the neutron polarization and $\mathbf{R}_i $ is the real-space position of atom $i$ in the unit cell, and the sum is over all atoms in the unit cell. Because $ \mu f(\mathbf{Q})$ is the quantity we are trying to evaluate, it can be taken as a variable in the following calculations. Assuming a magnetized ferromagnet, or a field-polarized paramagnet, the requirements for the measurement and ensuing calculation are to determine the flipping ratio, $R(\mathbf{Q}) = I^{++}/I^{--}$, obtained from polarized neutron diffraction measurements at a series of Bragg peaks with $\mathbf{P} \perp \mathbf{Q}$ and the moments parallel to $\mathbf{P}$ (where $I^{++}$ and $I^{--}$ are the two non-spin flip cross-section intensities). The flipping ratio also depends on the nuclear structure factor, $F_{\mathrm{N}}(\mathbf{Q})$, for the Bragg peaks measured, which can be calculated from the known crystal structure. The form factor can then be determined from the measured flipping ratio in terms of the nuclear and magnetic structure factors,

\begin{equation}\label{eq:FR}
R(\mathbf{Q}) = \frac{I^{++}}{I^{--}} = \frac{\left| F_N + F_M \right| ^2 }{\left| F_N - F_M \right| ^2}.
\end{equation}

When evaluating Eq.~\ref{eq:FR}, it is important to remember that the nuclear and magnetic structure factors are complex (i.e.\ $F_{\mathrm{N}}(\mathbf{Q}) = F_{\mathrm{N}}^{Re}(\mathbf{Q}) + iF_{\mathrm{N}}^{Im}(\mathbf{Q})$). For a more complete description of form factor measurements see, for example, Ch.~2 in Ref.~\cite{williams1988polarized}. 

\begin{table}[b]
\caption{Calculated flipping ratio values for various Bragg peaks assuming a single electron is evenly distributed between the V$_4$ cluster. The values calculated include just the structures factors (i.e.\ $f(\mathbf{Q})=1$ for all $\mathbf{Q}$).}
\label{tab:FRcub}
\begin{tabular}{|l|l|l|l|}
\hline
$H,K,L$ (r.l.u.) & $\left| F_N + F_M \right| ^2$ & $\left| F_N - F_M \right| ^2$ & $R(\mathbf{Q})$ \\ \hline
$1,1,1$          & 6.11        & 12.24       & 0.50       \\ \hline
$0,0,2$          & 4.84       & 2.82        & 1.71       \\ \hline
$2,2,0$          & 9.26       & 8.51        & 1.09       \\ \hline
$1,1,3$          & 9.15       & 6.72        & 1.36       \\ \hline
$2,2,2$          & 81.28       & 118.63      & 0.69       \\ \hline
$0,0,4$          & 41.03       & 20.25      & 2.03       \\ \hline
$3,3,1$          & 10.36       & 13.97      & 0.74       \\ \hline
$2,2,4$          & 11.12       & 9.19      & 1.21       \\ \hline
$3,3,3$          & 4.54       & 7.55      & 0.60       \\ \hline
$4,4,0$          & 143.85       & 106.29      & 1.35       \\ \hline
\end{tabular}
\end{table}

In Table S\ref{tab:FRcub}, we have calculated the theoretical values for $R(\mathbf{Q})$ assuming the cubic unit cell with the single electron evenly distributed between the V$_4$ cluster with $\mu = 0.25$ $\mu _{\mathrm{B}} /$V and $f(\mathbf{Q})=1$.  With no magnetic contribution the flipping ratio is of course unity, and here we note that the flipping ratios for the $(1,1,1)$-type peaks are opposite to most of the other peaks, which simply comes from the phase factors of the magnetic structure. We had hoped that this result would be qualitatively different if the moment resided solely on the apical V, but that turned out not to be the case for the peaks with a measurable magnetic signal.

\begin{figure}[t]
\includegraphics[scale=0.3]{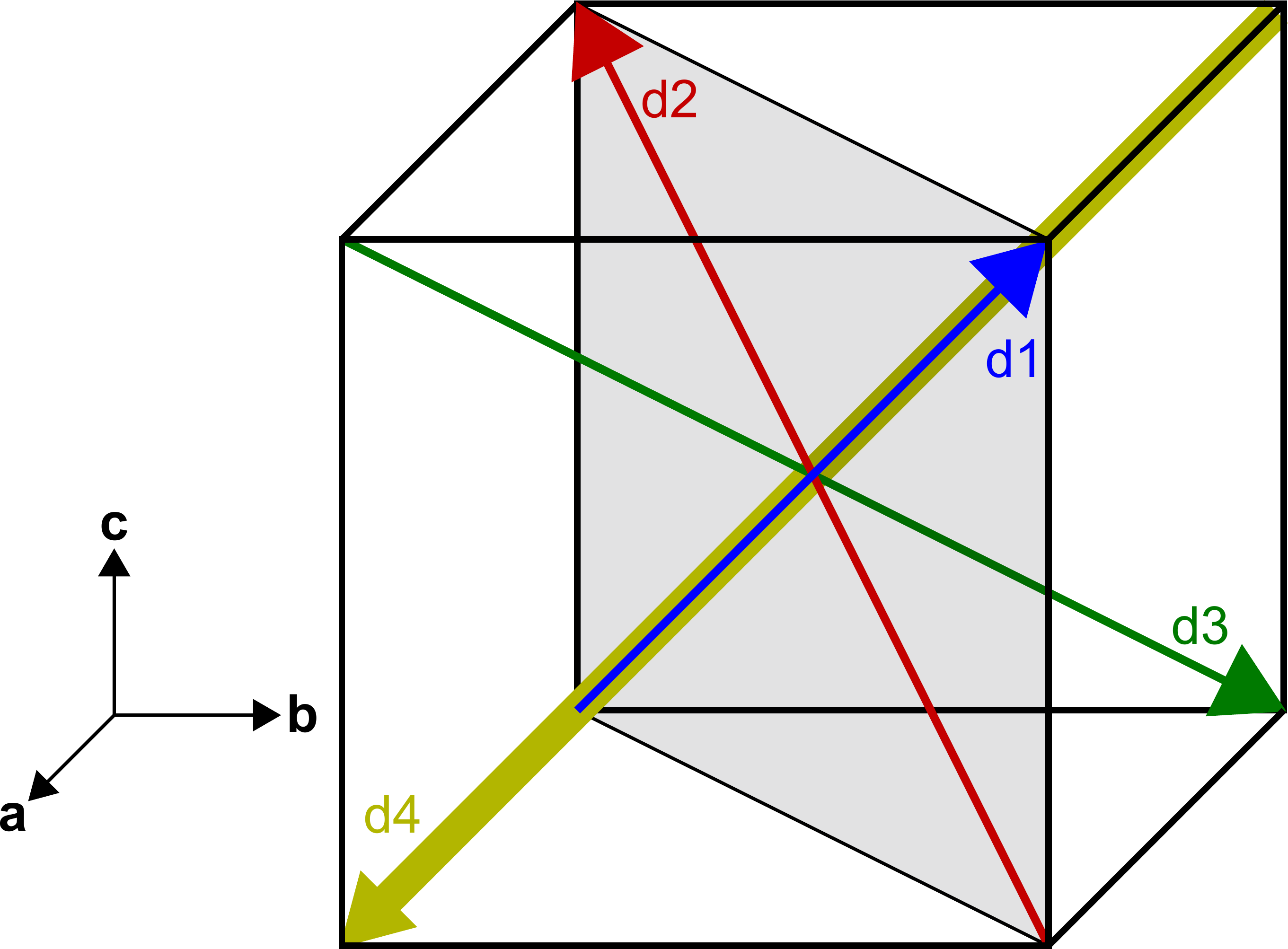}
\caption{Schematic of the orientations of each rhombohedral domain (d1--d4) with respect to the cubic structure. The cube outlined in black, and the crystallographic axes, represent the undistorted cubic crystal. The arrows show the different directions along which the cubic structure gets distorted. The direction of the arrows is also the direction that the spins align along for the different domains in the ferromagnetic phase. The gray plane shown is the scattering plane, $(H,H,L)$, which was used for the neutron experiments. \label{SFig2}}
\end{figure}

Going further, and to understand our $T=3$ K high instrumental resolution data, we calculated the flipping ratios at various reflections while taking into account the four rhombohedral crystallographic domains. A schematic showing the domains with respect to each other, and to the neutron scattering plane is shown in Fig.\ S\ref{SFig2}, where the arrows represent the four equivalent directions that the cubic unit cell can be stretched below the Jahn-Teller distortion temperature. The domains are labeled d1--d4, and the gray plane shown is the $(H, H, L)$ scattering plane in the cubic notation. The result of the domains in reciprocal space is that some cubic peaks split along $Q$ due to different reflections from different domains cutting into the scattering plane. 

\begin{table*}[t]
\caption{Calculated flipping ratio values, $R(Q)$, for various Bragg peaks for three models: (1) assuming a single electron is evenly distributed between the V$_4$ cluster ($0.25$ $\mu _B /$V), (2) the inhomogeneous distribution of the electron from DFT calculations, where the values have been scaled so that the total moment adds up to $1$ $\mu _B$ ($0.204$ $\mu _B /$basal V; $0.387$ $\mu _B /$apical V), and (3) the electron is located only on the apical V ($1$ $\mu _B /$apical V). This table differs from Table S\ref{tab:FRcub}, in that the four crystallographic domains were considered. The first column lists the pseudo-cubic $(H,K,L)$, and the second column lists the rhombohedral domains which contribute intensity around that point. The domains are grouped by equivalent $Q$'s. For example, the cubic $(1,1,1)$ splits into two peaks. The lower $Q$ peak is only due to the $(1,1,1)$ peak from d1, but the higher $Q$ peak is a superposition of d2, d3, and d4 Bragg peaks. The values calculated include just the structures factors (i.e.\ $f(\mathbf{Q})=1$ for all $\mathbf{Q}$).}
\label{tab:FRrhom}
\begin{tabular}{|>{\raggedright}p{1.5cm}|l|>{\raggedright}p{2.5cm}|>{\raggedright}p{2.5cm}|>{\raggedright}p{2.5cm}|}
\hline
`cubic' $H,K,L$					 			& rhombohedral domain and corresponding $H, K, L$								& $R(Q)$ ($0.25$ $\mu _B /$V)	& $R(Q)$ ($0.204$ $\mu _B /$basal V; $0.387$ $\mu _B /$apical V) 	& $R(Q)$ ($1$ $\mu _B /$apical V) \tabularnewline \hline
\multirow{2}{*}{$1,1,1$}      & d1 $(1, 1, 1)$                																& 0.51												&0.65																															& 1.80  \tabularnewline \cline{2-5} 
                              & d2 $(0, 0, -1)$, d3 $(0,-1,0)$), d4 $(-1, 0, 0)$        			& 0.50												&0.47																															& 0.35  \tabularnewline \hline
\multirow{1}{*}{$0,0,2$}      & d1 $(1,1,0)$, d2 $(1,1,0)$, d3 $(-1,-1,0)$, d4 $(-1,-1,0)$    & 1.74												&1.69																															& 1.39  \tabularnewline \hline 
\multirow{2}{*}{$2,2,0$}      & d1 $(1,1,2)$, d2 $(-1,-1,-2)$            											& 1.08												&0.84																															& 0.28  \tabularnewline \cline{2-5} 
                              & d3 $(1,-1,0)$, d4 $(-1,1,0)$            											& 1.10												&1.41																															& 4.58  \tabularnewline \hline
\multirow{3}{*}{$1,1,3$}      & d1 $(2,2,1)$               																		& 1.30												&1.60																															&	4.16	\tabularnewline \cline{2-5} 
                              & d3 $(-1,-2,0)$, d4 $(-2,-1,0)$            										& 1.40												&1.45																															&	1.63	\tabularnewline \cline{2-5}
															& d2 $(1,1,-1)$               																	& 1.36												&1.07																															&	0.37	\tabularnewline \hline 
\multirow{2}{*}{$2,2,2$}      & d1 $(2,2,2)$                																	& 0.67												&0.67																															&	0.66	\tabularnewline \cline{2-5} 
                              & d2 $(0,0,-2)$, d3 $(0,-2,0)$, d4 $(-2,0,0)$        						& 0.69												&0.69																															&	0.69	\tabularnewline \hline
\multirow{1}{*}{$0,0,4$}      & d1 $(2,2,0)$, d2 $(2,2,0)$, d3 $(-2,-2,0)$, d4 $(-2,-2,0)$    & 2.03												&2.03																															&	2.03	\tabularnewline \hline
\multirow{3}{*}{$3,3,1$}      & d1 $(2,2,3)$                																	& 0.72												&0.72																															&	0.73	\tabularnewline \cline{2-5} 
                              & d2 $(-1,-1,-3)$                																& 0.75												&1.00																															&	3.90	\tabularnewline \cline{2-5}
															& d3 $(1,-2,0)$, d4 $(-2,1,0)$            											&	0.75												&0.65																															&	0.32	\tabularnewline \hline
\multirow{3}{*}{$2,2,4$}      & d1 $(3,3,2)$               																		& 1.29												&1.53																															&	3.48	\tabularnewline \cline{2-5}
															&	d3 $(-1,-3,0)$, d4 $(-3,-1,0)$																&	1.21												&1.00																															&	0.44	\tabularnewline \cline{2-5}
                              & d2 $(1,1,-2)$                																	& 1.20												&1.46																															&	3.54	\tabularnewline \hline
\multirow{2}{*}{$3,3,3$}      & d1 $(3,3,3)$               																		& 0.69												&0.55																															&	0.18	\tabularnewline \cline{2-5} 
                              & d2 $(0,0,-3)$, d3 $(0,-3,0)$, d4 $(-3,0,0)$       						& 0.65												&0.71																															&	1.06	\tabularnewline \hline
\multirow{2}{*}{$4,4,0$}      & d1 $(2,2,4)$, d2 $(-2,-2,-4)$            											& 1.37												&1.35																															&	1.28	\tabularnewline \cline{2-5} 
                              & d3 $(2,-2,0)$, d4 $(-2,2,0)$           												& 1.34												&1.37																															&	1.47	\tabularnewline \hline																																																												
\end{tabular}
\end{table*}

The flipping ratio calculations for various Bragg reflections are presented in Table S\ref{tab:FRrhom}, where the first column is a cubic $(H,K,L)$ reflection and the second column contains the corresponding rhombohedral $(H,K,L)$ reflections which contribute intensity about the cubic point. The rows are split according to $Q$ value. The third column contains the flipping ratio values assuming all moments are field polarized along a vertically applied magnetic field with the unpaired electron within a V$_4$ cluster evenly distributed. This is the same assumption as the calculations presented in Table S\ref{tab:FRcub}. One can see the values are almost exactly the same as those assuming a cubic notation, further validating our use of the cubic unit cell for calculations presented in the main text. The fourth column presents calculations assuming the electron is distributed inhomogeneously between the basal and apical V atoms according to the DFT results, which predicted $0.221$ $\mu _B$/basal V and $0.419$ $\mu _B$/apical V. The calculations in Table S\ref{tab:FRrhom} used this distribution but scaled the total moment to be $1$ $\mu _B$ in order to compare with the other models. The last column presents calculations assuming the electron is localized to the apical V, with a value of $1$ $\mu _{\mathrm{B}}$. Neither the equal distribution or apical assumption can explain the flipping ratio values measured with high resolution, like those in Fig.\ 5(c) and (d) of the main text. In those figures, the lower-$Q$ peak is only due to intensity coming from d1, and has a flipping ratio of $\approx 1$. This would be the case if the moments were not polarized along the applied field direction, but still along---or mostly along---the local $[1,1,1]$-like directions due to the uniaxial anisotropy. Evaluating the rest of the data with this assumption proved difficult with the high-resolution data, however, because of crystallite formation below the Jahn-Teller distortion as we discuss shortly.

The main text presented the results of only the 4 V model (with equal electron distribution) and apical only model. The differences in the flipping ratios between the DFT inhomogeneous electron distribution and the 4 V model are much smaller than between the 4 V model and apical only model. This is especially true as the moment size is reduced from saturation, as is the case at 20 K with a 2 T field. Fig.\ 3(a) of the main text shows the moment per V to be $\approx 0.1$ $\mu _B$ at 20 K and 2 T, and to illustrate the differences between the models with this total moment value, we present the corresponding flipping ratio calculations in Table S\ref{tab:FRrhom_smmu}.

\begin{table*}[t]
\caption{Calculated flipping ratio values, $R(Q)$, for various Bragg peaks for three models, where the total moment used ($0.4$ $\mu _B$) matches that extracted from the 20 K data at 2 T from Fig.\ 3(a) of the main text. The three models are: (1) assuming the moment is evenly distributed between the V$_4$ cluster ($0.1$ $\mu _B /$V), (2) the inhomogeneous distribution of the moment from DFT calculations, where the values have been scaled so that the total moment adds up to $0.4$ $\mu _B$ ($0.082$ $\mu _B /$basal V; $0.155$ $\mu _B /$apical V), and (3) the moment is located only on the apical V ($0.4$ $\mu _B /$apical V). This table differs from Table S\ref{tab:FRcub}, in that the four crystallographic domains were considered. The first column lists the pseudo-cubic $(H,K,L)$, and the second column lists the rhombohedral domains which contribute intensity around that point. The domains are grouped by equivalent $Q$'s. For example, the cubic $(1,1,1)$ splits into two peaks. The lower $Q$ peak is only due to the $(1,1,1)$ peak from d1, but the higher $Q$ peak is a superposition of d2, d3, and d4 Bragg peaks. The values calculated include just the structures factors (i.e.\ $f(\mathbf{Q})=1$ for all $\mathbf{Q}$).}
\label{tab:FRrhom_smmu}
\begin{tabular}{|>{\raggedright}p{1.5cm}|l|>{\raggedright}p{2.5cm}|>{\raggedright}p{2.5cm}|>{\raggedright}p{2.5cm}|}
\hline
`cubic' $H,K,L$					 			& rhombohedral domain and corresponding $H, K, L$								& $R(Q)$ ($0.1$ $\mu _B /$V)	& $R(Q)$ ($0.082$ $\mu _B /$basal V; $0.155$ $\mu _B /$apical V) 	& $R(Q)$ ($0.4$ $\mu _B /$apical V) \tabularnewline \hline
\multirow{2}{*}{$1,1,1$}      & d1 $(1, 1, 1)$                																& 0.76												&	0.84																														& 1.29  \tabularnewline \cline{2-5} 
                              & d2 $(0, 0, -1)$, d3 $(0,-1,0)$), d4 $(-1, 0, 0)$        			& 0.76												&	0.74																														& 0.65  \tabularnewline \hline
\multirow{1}{*}{$0,0,2$}      & d1 $(1,1,0)$, d2 $(1,1,0)$, d3 $(-1,-1,0)$, d4 $(-1,-1,0)$    & 1.25												&	1.24																														& 1.18  \tabularnewline \hline 
\multirow{2}{*}{$2,2,0$}      & d1 $(1,1,2)$, d2 $(-1,-1,-2)$            											& 1.03												&	0.93																														& 0.60  \tabularnewline \cline{2-5} 
                              & d3 $(1,-1,0)$, d4 $(-1,1,0)$            											& 1.04												&	1.15																														& 1.80  \tabularnewline \hline
\multirow{3}{*}{$1,1,3$}      & d1 $(2,2,1)$               																		& 1.11												&	1.21																														&	1.76	\tabularnewline \cline{2-5} 
                              & d3 $(-1,-2,0)$, d4 $(-2,-1,0)$            										& 1.15												&	1.16																														&	1.24	\tabularnewline \cline{2-5}
															& d2 $(1,1,-1)$               																	& 1.13												&	1.03																														&	0.66	\tabularnewline \hline 
\multirow{2}{*}{$2,2,2$}      & d1 $(2,2,2)$                																	& 0.85												&	0.85																														&	0.85	\tabularnewline \cline{2-5} 
                              & d2 $(0,0,-2)$, d3 $(0,-2,0)$, d4 $(-2,0,0)$        						& 0.86												&	0.86																														&	0.86	\tabularnewline \hline
\multirow{1}{*}{$0,0,4$}      & d1 $(2,2,0)$, d2 $(2,2,0)$, d3 $(-2,-2,0)$, d4 $(-2,-2,0)$    & 1.32												&	1.32																														&	1.33	\tabularnewline \hline
\multirow{3}{*}{$3,3,1$}      & d1 $(2,2,3)$                																	& 0.87												&	0.87																														&	0.87	\tabularnewline \cline{2-5} 
                              & d2 $(-1,-1,-3)$                																& 0.89												&	1.00																														&	1.70	\tabularnewline \cline{2-5}
															& d3 $(1,-2,0)$, d4 $(-2,1,0)$            											&	0.89												&	0.84																														&	0.64	\tabularnewline \hline
\multirow{3}{*}{$2,2,4$}      & d1 $(3,3,2)$               																		& 1.11												&	1.19																														&	1.63	\tabularnewline \cline{2-5}
															&	d3 $(-1,-3,0)$, d4 $(-3,-1,0)$																&	1.08												&	1.00																														&	0.71	\tabularnewline \cline{2-5}
                              & d2 $(1,1,-2)$                																	& 1.08												&	1.17																														&	1.65	\tabularnewline \hline
\multirow{2}{*}{$3,3,3$}      & d1 $(3,3,3)$               																		& 0.85												&	0.78																														&	0.52	\tabularnewline \cline{2-5} 
                              & d2 $(0,0,-3)$, d3 $(0,-3,0)$, d4 $(-3,0,0)$       						& 0.83												&	0.86																														&	1.03	\tabularnewline \hline
\multirow{2}{*}{$4,4,0$}      & d1 $(2,2,4)$, d2 $(-2,-2,-4)$            											& 1.13												&	1.13																														&	1.11	\tabularnewline \cline{2-5} 
                              & d3 $(2,-2,0)$, d4 $(-2,2,0)$           												& 1.12												&	1.13																														&	1.17	\tabularnewline \hline																																																												
\end{tabular}
\end{table*}

\section{Crystallite Formation}

\begin{figure}[b]
\includegraphics[scale=0.15]{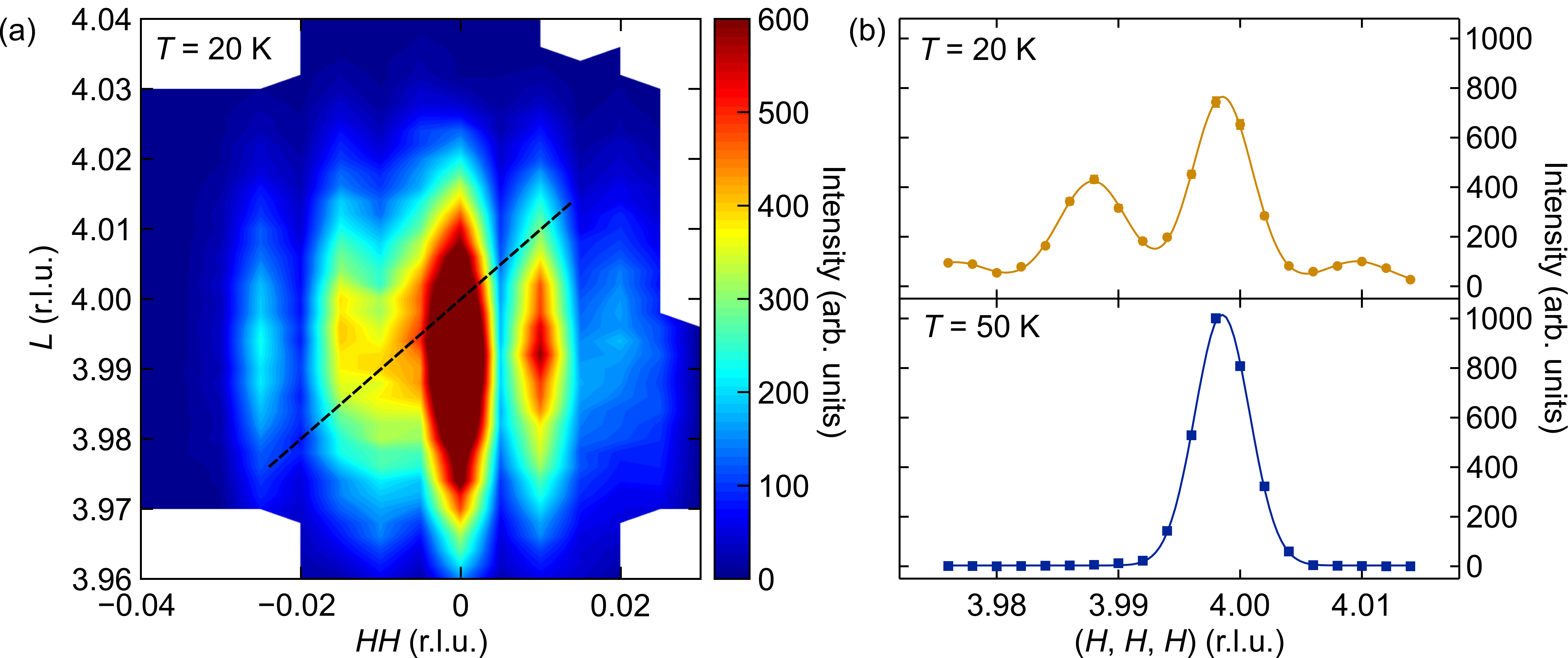}
\caption{(a) Map of the $(H, H, L)$ plane about the cubic $(0, 0, 4)$ peak at $T=20$ K taken with the high instrumental resolution. Splitting of the peak is along the mosaic direction, such that all peaks have the same $Q$. The dashed line shows a cut along the $(H, H, H)$ direction, and this cut is shown in panel (b) at two different temperatures. The top shows the same temperature, $T=20$ K, as in (a). This is below the Jahn-Teller transition temperature where four crystallographic domains are expected. The bottom panel shows the same $(H, H, H)$ cut at $T=50$ K. The single peak shows that the crystallite formation beneath the Jahn-Teller distortion is reversible.\label{SFig3}}
\end{figure} 

Four crystallographic domains form below the Jahn-Teller distortion, but we found that more than four crystallites are actually forming. The orientation of each crystallite with respect to one another leads to multiple Bragg peaks along the mosaic direction in reciprocal space. Fig.\ S\ref{SFig3}(a) shows an $(H,H,L)$ map about the $(0,0,4)$ point in cubic notation at $T=20$ K. It is expected that contributions to the Bragg scattering from the four different domains should all coincide at a single point in reciprocal space here, however multiple peaks are seen spanning the mosaic direction, meaning they all have the same $Q$ value. When taking a cut along the $(H,H,H)$ direction, as we would when collecting flipping ratio data, the multiple peaks are clearly seen, as shown in the top panel of Fig.\ S\ref{SFig3}(b). The sample quality, however, is not to blame, as the multiple crystallites form a single crystallite when the temperature is raised above that of the Jahn-Teller distortion, as seen in the bottom panel of Fig.\ S\ref{SFig3}(b) at $T=50$ K. This reversible process was seen at other Bragg peak positions. We were not able to reliably assign which domain contributed to which peak for all peaks seen. The course resolution used for the $T=15$ K and $T=20$ K data integrated over all domains and crystallites, avoiding this problem. Course resolution data were taken for $T=3$ K, but the analysis was inconclusive, which is why we initially attempted the high resolution measurements. It was then that we found the moments could not be fully field polarized with 2 T at $T = 3$ K. 

\section{Analysis of density functional theory results}

\begin{figure}[t]
\includegraphics[width=0.49\textwidth]{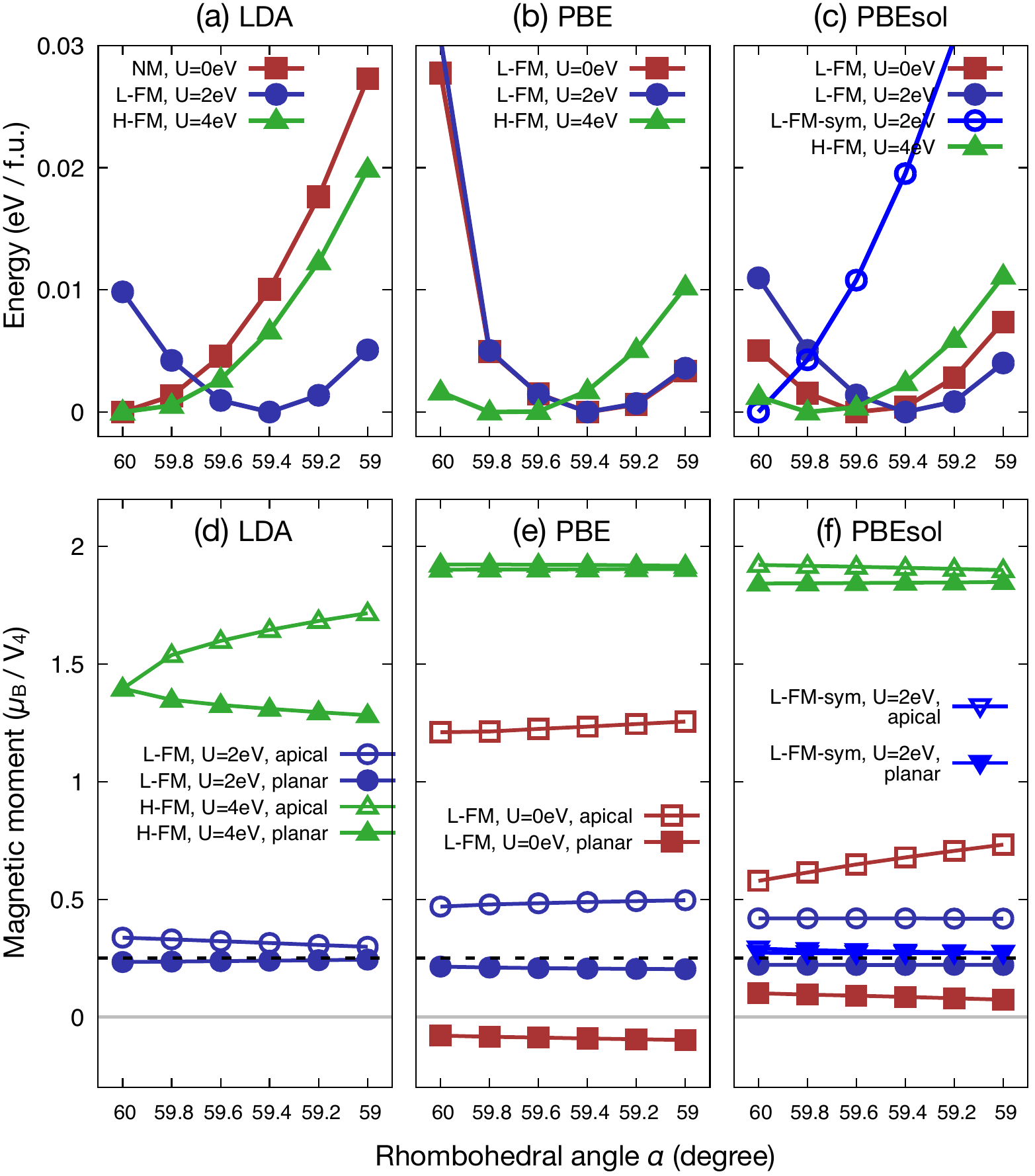}
\caption{
Plots of (a-c) total energy and (d-f) magnetic moment values as a function of rhombohedral angle $\alpha$ at different $U_{\rm eff}$, where spin moments of apical and planar V sites are separately shown in (d-f). NM, L-FM, and H-FM stand for nonmagnetic, low-spin ferromagnetic, and high-spin ferromagnetic phases, respectively. 
\label{SFig_EnM}}
\end{figure}

To understand the low-temperature ferromagnetism in GaV$_4$S$_8$ and its behavior in the presence of electron correlations, density functional theory (DFT) calculations were carried out with the inclusion of on-site Coulomb repulsion incorporated via a simplified rotationally-invariant DFT+$U_{\rm eff}$ method~\cite{Dudarev1998}. For the test of how GaV$_4$S$_8$ behaves under different choices of exchange-correlation functionals, we employed the Ceperley-Alder parametrization of the local-density approximation (LDA), \cite{LDA} the Perdew-Burke-Ernzerhof generalized-gradient approximation (PBE), \cite{PBE} and the variant of PBE for crystalline solids (PBEsol). \cite{PBEsol} For the choice of DFT code and computational parameters, refer to the Methods section in the main text. The unit cell volume at ambient conditions was optimized for each choice of exchange-correlation functional and $U_{\rm eff}$ value, after which the rhombohedral angle $\alpha$ was varied at fixed volume. A collinear ferromagnetic initial condition was chosen for all calculations, and it was checked that an intra-V$_4$-cluster antiparallel arrangement of $V$ spin moments could only be stabilized in the H-FM (high-spin configuration) phases. 

\begin{figure}[t]
\includegraphics[width=0.48\textwidth]{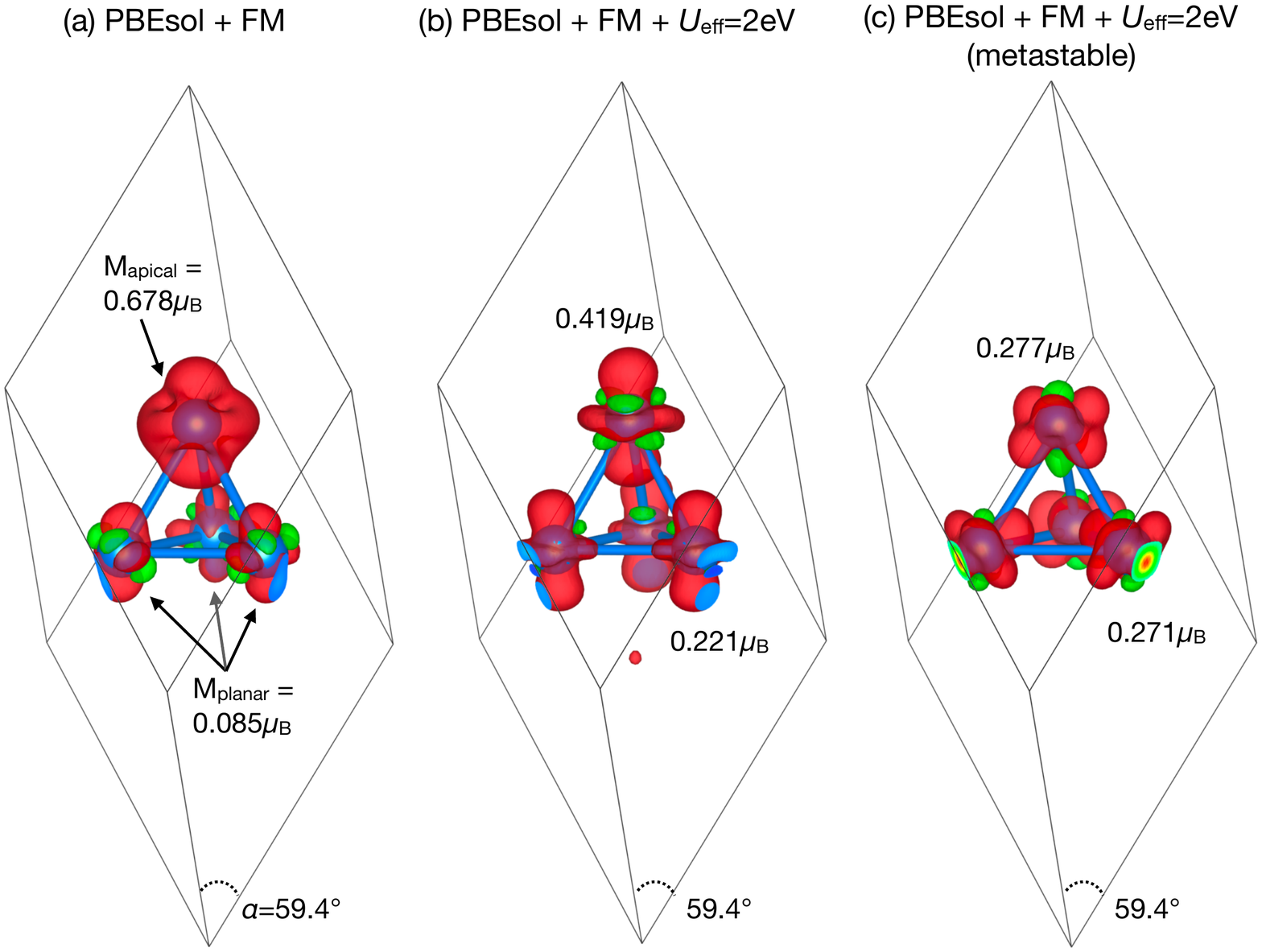}
\caption{
Spin density isosurfaces of GaV$_4$S$_8$ from PBEsol+$U_{\rm eff}$ density functional calculations with different $U_{\rm eff}$ values: (a) at $U_{\rm eff}$ = 0 eV and (b,c) $U_{\rm eff}$ = 2 eV, where (b) and (c) are from the ground and metastable configurations respectively. Spin densities are plotted by employing a rhombohedral angle $\alpha$ = 59.4$^\circ$, and black lines depict the primitive unit cell. 
\label{SFig_spin}}
\end{figure}

Fig.~S\ref{SFig_EnM} and Table S\ref{tab:DFTdata} summarize the DFT+$U_{\rm eff}$ results, showing the behavior of the total energy and magnetism as a function of $\alpha$. Therein NM, L-FM, and H-FM stand for nonmagnetic, $S = 1/2$ low-spin, and $S = 5/2$ (LDA) or $7/2$ (PBE/PBEsol) high-spin configurations as discussed in Ref.~\onlinecite{HSK2018}. Several features of computational results are noteworthy:
\begin{itemize}
\item The magnetism and rhombohedral distortion are strongly coupled to each other. In particular, the experimentally found rhombohedral distortion of $\alpha \approx$ 59.6$^\circ$~\cite{Pocha2000} is only reproduced in the L-FM configuration with small $U_{\rm eff}$ of  $\leq 2$ eV. This is because (i) the bulk dipole moment cannot be stabilized in the metallic NM phase and (ii) the H-FM phases do not have strong Jahn-Teller instabilities like the L-FM case, due to the closed-shell orbital configuration and weaker $V_4$ clustering.~\cite{HSK2018}
\item Employing no $U_{\rm eff}$, except for the LDA case, results in a strongly localized spin moment distribution at the apical V site (denoted as Phase 2 hereafter and in Table S\ref{tab:DFTdata}), which remains stable as $U_{\rm eff}$ is included. This localization of the spin moment seems to emerge from the momentum space dispersion of the partially filled T$^2$ molecular orbital triplet band. Interestingly, introducing $U_{\rm eff}$ facilitates the convergence of phases with more equally distributed spin density over the 4 V sites in the V$_4$ cluster (denoted as Phase 1), while phase 2 is lower in energy over the whole $U_{\rm eff}$-range in the PBEsol case (see Table S\ref{tab:DFTdata}). The spin localization is strongest in the PBE functional results, and weakest in the LDA case. 
\item Table S\ref{tab:DFTdata} summarizes the computed structural parameters and V-site magnetizations obtained from two different local minima states (Phase 1 and 2). \footnote{Note that an enhanced plane-wave energy cutoff of 500 eV was employed to generate data shown in this table for more accurate total energy comparison, which yields slightly different sizes of vanadium spin moments compared to shown in Fig.~S\ref{SFig_spin}.} Note that the presence of more than one local minima state in GaV$_4$S$_8$ is also noticed in a recent DFT+$U$ computational study. \cite{Wang2019} Phase 1 in Table S\ref{tab:DFTdata} shows spin moments which are close to equi-distributed over all V sites, while the spin moments starting from phase 2 are mostly localized at the apical V site. Phase 2 corresponds to the data presented in Ref.~\onlinecite{Wang2019} and, as mentioned therein, is lower in energy than phase 1 by 0.1 $\sim$ 0.3 eV per formula unit depending on the value of $U_{\rm eff}$. Phase 1, on the other hand, is more consistent with the measured spin distribution in this work, and it also captures better agreements of the rhombohedral angle $\alpha$ and the degree of V$_4$ elongation ($d$V-V$_{\rm ap}$ / $d$V-V$_{\rm pl}$) between theory and experiments around $U_{\rm eff}$ = 1 eV. Hence we focus on the phase 1 results in this study because of its better agreement of physically important parameters (spin distribution, rhombohedral angle, and the degree of V$_4$ cluster elongation) with experiments. 
\item The results of the calculations are strongly dependent on the initial conditions, suggesting multiple local-minima states in this system. Specifically in the PBEsol functional results, three configurations with different spin moment distributions can be obtained around $U_{\rm eff}$ = 2 eV; even within phase 1, two different patterns of the spin distribution occur (Figs.~\ref{SFig_EnM}(c) and (f), and Figs.~\ref{SFig_spin}(b) and (c)). The case with equally distributed spin moments, while higher in energy, favors a cubic structure, while the ground state configuration with an imbalanced spin moment distribution favors a rhombohedrally distorted phase. Their total energies versus the rhombohedral angle are shown in Figs.~\ref{SFig_EnM}(c) and (f).
\end{itemize}

\begin{table*}[b]
\caption{Summary of PBEsol+$U_{\rm eff}$ calculation results in the low-temperature rhombohedral phase, in comparison with experimentally measured structural parameters from Ref.~\onlinecite{Pocha2000} and \onlinecite{Powell2007}. Below  $d{\rm V-V}_{\rm ap}$ and  $d{\rm V-V}_{\rm pl}$ are the longer and shorter intra-cluster V-V distances, respectively. $M-{\rm V_{ap}}$ and $M-{\rm V_{pl}}$ denote magnetization at the apical and planar V-sites in the $V_4$ cluster. Total energies per formula unit of phase 1 and 2 from VASP calculations, $E_1$ and $E_2$ respectively, and their difference at the same $U_{\rm eff}$ value are shown also.}
\label{tab:DFTdata}
\begin{tabular}{p{22mm}p{14mm}p{14mm}p{14mm}p{14mm}p{14mm}p{14mm}p{14mm}p{14mm}p{14mm}p{14mm}}
Phase 1 &&&&&&&&&& \\
$U_{\rm eff}$ \newline (eV) & $a$ \newline (\AA) & $\alpha$ \newline (deg.) & $d{\rm V-V}_{\rm ap}$  \newline (\AA) & $d{\rm V-V}_{\rm pl}$ \newline (\AA) & $\frac{d{\rm V-V}_{\rm ap}}{d{\rm V-V}_{\rm pl}}$ & 
$M-{\rm V}_{\rm ap}$ \newline ($\mu_{\rm B}$) & $M-{\rm V}_{\rm pl}$ \newline ($\mu_{\rm B}$) & Gap size \newline (eV) & $E$ \newline (eV/f.u.) & $E_1 - E_2$ \newline (eV/f.u.) \\ \hline
1.0 & 6.780&	59.500&	2.865&	2.746&	1.043&	0.249&	0.252&	0.270&	-85.408&	0.116\\
1.5&	6.792&	59.431&	2.881&	2.743&	1.050&	0.361&	0.231&	0.375&	-83.342&	0.224\\
2.0&	6.804&	59.383&	2.899&	2.742&	1.057&	0.426&	0.220&	0.569&	-81.297&	0.356\\
&&&&&&&&&& \\
Phase 2 &&&&&&&&&& \\ \hline
0&	6.765&	59.534&	2.853&	2.749&	1.038&	0.686&	0.082&	0.000&	-89.615&\\	
0.5&	6.791&	59.354&	2.893&	2.752&	1.051&	1.200&	-0.083&	0.110&	-87.538&\\	
1.0&	6.813&	59.263&	2.925&	2.763&	1.059&	1.505&	-0.189&	0.167&	-85.524&\\
1.5&	6.832&	59.199&	2.951&	2.776&	1.063&	1.710&	-0.271&	0.239&	-83.566&\\	
2.0&	6.851&	59.140&	2.975&	2.789&	1.067&	1.875&	-0.346&	0.309&	-81.653&\\
&&&&&&&&&& \\
Exp. results &&&&&&&&&& \\ \hline
Pocha et al. \newline LT, 20K &	6.834&	59.660&	2.898&	2.826&	1.025&&&&& \\
Powell et al. \newline LT, 20K & 	6.839&	59.643&	2.943&	2.856&	1.030&&&&& \\
Powell et al. \newline LT, 4.2K &	6.840&	59.616&	2.920&	2.846&	1.026&&&&&
\end{tabular}
\end{table*}

The presence of multiple local minima states in the low-temperature phase of GaV$_4$S$_8$ implies a complicated interplay between the charge, orbital, spin, and lattice degrees of freedom in this system. Further, it was suggested recently that inter-site correlation effects may also be crucial in understanding several structural properties of this compound even in the high-temperature cubic phase. \cite{HSK2018} Hence more elaborate computational studies of GaV$_4$S$_8$ beyond the DFT+$U$ level should follow for a better understanding of the ground state and multiferroic properties.

\bibliography{supplement}